\documentclass[epj,nopacs]{svjour}
\usepackage{graphics}
\usepackage{latexsym,graphicx,epsfig,psfrag,here,amsmath}
\allowdisplaybreaks
\usepackage{amsmath} 
\usepackage{longtable}
\usepackage{amssymb} 
\usepackage{bbm}
\usepackage{cite}  

\def\bra#1{\left\langle #1\right|}

\def\ket#1{\left| #1\right\rangle}

\def\braket#1{\left\langle  #1\right\rangle}

\begin{document}
%

\title{Form factors in the radiative pion decay}
\author{Vicent Mateu 
\and Jorge Portol\'es 
}                     
%
%
\institute{IFIC, Universitat de Val\`encia -- CSIC, Apt. Correus 22085, E-46071
Val\`encia, Spain}
\date{}
%
\abstract{
We perform an analysis of the form factors that rule the structure-dependent amplitude
in the radiative pion decay.
The resonance contributions to $\pi \rightarrow e \, \nu_{e} \gamma$ 
decays are computed through the proper construction of the vector and axial-vector form factors by setting the QCD driven asymptotic properties of the three-point Green functions $\langle VVP \rangle$ and $\langle VAP \rangle$,
and by demanding the smoothing of the form factors at high transfer of momentum. A comparison between
theoretical and experimental determinations of the form factors is also carried out. We also consider and evaluate the 
role played by a non-standard tensor form factor. We conclude that, at present and due to the hadronic incertitudes, the search for New Physics in this process is not feasible.
} 
\maketitle
%

\section{Introduction}
The radiative decay of the pion is a suitable process to be analysed within Chiral Perturbation
Theory ($\chi$PT) \cite{Weinberg:1978kz,Holstein:1986uj}, the effective field theory
of QCD in the very low-energy region. This framework provides the structure of the relevant
form factors through~: i) a polynomial expansion in momenta, essentially driven by the contributions
of heavier degrees of freedom that have been integrated out, and ii) the required chiral logarithms
generated by the loop expansion and compelled by unitarity. 
Both contributions correspond to the chiral expansion in 
$p^2/M_V^2$ and $p^2/\Lambda_{\chi}^2$, respectively, where $M_V$ is the mass of the lightest vector
resonance, $\Lambda_{\chi} \sim 4 \pi F$ and $F$ is the decay constant of the pion. 
Hence their magnitude is, in principle, comparable. The chiral logarithms have thoroughly been studied in 
later years up to ${\cal O}(p^6)$ both in $SU(2)$ \cite{Bijnens:1996wm} and $SU(3)$ \cite{Ametller:1993hg,Geng:2003mt}.
\par
However the size of the polynomial contributions is more controversial. They involve short-distance 
dynamics through the chiral low-energy constants (LECs) of the $\chi$PT Lagrangian and their determination
from QCD is a difficult non-perturbative problem. Phenomenology and theoretical arguments suggest 
that the main role is played by the physics at the scale $M_V$, {\it i.e.} the physics of low-lying
resonances, and a lot of effort has been dedicated to pursue
this goal \cite{Ecker:1988te,Cirigliano:2006hb}. This assumption, widely known as resonance saturation 
of the LECs in $\chi$PT, implies that the structure of the form factors is given by the pole dynamics
of resonances and this hint works well in all known cases in the allowed range of energies as, {\it e.g.}
scalar, vector and axial-vector form factors in hadronic decays of the tau lepton 
\cite{Kuhn:1990ad,GomezDumm:2003ku}.
\par
Because $M_{\pi} \ll M_V$, it is expected that the structure of the form factors in pion decays
should be less relevant and, accordingly, the approach provided by $\chi$PT should be good enough even
when only the first terms of the chiral expansion are included. This is the case of the radiative pion
decay, namely $\pi \rightarrow \ell \, \nu_{\ell} \gamma$, $\ell= \mu, e$, where constant form factors (that correspond to the 
leading order contribution in the chiral expansion) have widely been employed
in the analyses of data. However the PIBETA collaboration \cite{Frlez:2003pe} showed that a strong 
discrepancy between theory and experiment arises for the branching ratio of the process in a specific
region of the electron and photon energies. Lately the same collaboration, after their 2004 analysis,
concludes that the discrepancy has faded away \cite{Pokanic06}. Curiously enough this decay
has a persistent story of deceptive comparisons between theory and experiment \cite{Bolotov:1990yq}
that have prompted the publication of proposals beyond the Standard Model (SM) to account
for the variance \cite{Poblaguev:1990tv,Belyaev:1991gs,Herczeg:1994ur,Poblaguev:2003ib,Chizhov:2004tu}.
Between these it has received particular attention the possibility of allowing a tensor contribution
that could explain the discrepancy by interfering destructively with the Standard Model prescription
though showing some inconsistency with the corresponding tensor contribution in nuclear $\beta$ decay
\cite{Quin:1993vh}. Related with this issue it is essential, seeking to discern the presence
of a new physics contribution to the radiative decay of the pion, to provide an accurate profile
of the involved form factors within QCD.
\par
In order to settle the Standard Model description of the vector and axial-vector form factors
participating in $\pi \rightarrow \ell \, \nu_{\ell} \gamma$ decays we study, in this article, the structure
provided by the lightest meson resonances. This is very much relevant on the experimental side because
high-statistics experiments as PIBETA \cite{Pokanic06} already are able to determine, for instance, 
the slopes of the form factors involved in these decays. The procedure consists, essentially, 
in the construction of the relevant three-point Green functions (GF) in the 
resonance energy region with a finite spectrum of states at the poles of the corresponding
meromorphic functions. This can be carried out by employing a parametric {\em ansatz} or a Lagrangian
theory (we work it out in both cases). In a second stage several constraints are imposed on the
parameters or coupling constants, namely chiral symmetry at $q^2 \ll M_{R}^2$ and the asymptotic
behavior ruled by the Operator Product Expansion of the GF for $q^2 \gg M_{R}^2$, where $M_R$ are
short for typical masses of the resonances in the poles and $q^2$ an indicative squared transfer
of momenta of this energy region. This method, guided by the large-$N_C$ limit of QCD
\cite{'tHooft:1973jz} ($N_C$ is short for the number of colours), has proven to be efficient 
\cite{Peris:1998nj,Knecht:2001xc,Ruiz-Femenia:2003hm,Bijnens:2003rc,Cirigliano:2004ue,Ecker:1989yg,otherGF} in 
order to collect the constraints that drive form factors of QCD currents.
\par
Section~2 is devoted to the construction of the vector
and axial-vector form factors using the procedure outlined above. The study of 
the beyond the SM contribution of a tensor current to the radiative pion decay is performed in 
Section~3 and, finally, Section~4 describes the analysis of the photon spectrum in this process.
After the Conclusions in Section~5, three appendices complete the setting of this article.

\section{Radiative pion decay~: Vector and axial-vector form-factors}
The amplitude that describes the $\pi^+ \rightarrow \ell^+ \nu_{\ell} \gamma$ process can be split
into two different contributions~:
\begin{equation}
 M(\pi^+ \rightarrow \ell^+ \nu_{\ell} \gamma) \, = \, M_{IB} \, + \, M_{SD} \; .
\end{equation}
Here $M_{IB}$ is the inner bremsstrahlung (IB) amplitude where the photon is radiated by the 
electrically charged external legs, either pion or lepton; consequently the interaction is driven by
the axial-vector current. $M_{SD}$ is the structure-dependent (SD) contribution where the photon
is emitted from intermediate states generated by strong interactions. In this later case
both vector and axial-vector form factors arise from the hadronization of the QCD currents within the
Standard Model.
\par
Because $\pi^+ \rightarrow e^+ \nu_{e}$ is helicity suppressed, the IB contribution to its radiative
counterpart suffers the same inhibition and, consequently, the electron case is the appropriate 
channel to uncover the non-perturbative SD amplitude. Contrarily, the
$\pi^+ \rightarrow \mu^+ \nu_{\mu} \gamma$ decay is fairly dominated by IB. As a consequence
the $\pi^+ \rightarrow e^+ \nu_e \gamma$ is of great interest to investigate the hadronization of
the currents contributing to the SD amplitudes that are driven, within the Standard Model, by the vector 
($F_V(q^2)$) and axial-vector ($F_A(q^2)$) form factors defined by
\footnote{We use the convention $\varepsilon_{0123}=+1$ for the Levi-Civita tensor $\varepsilon_{\mu\nu\alpha\beta}$ throughout this paper.}~:
 \begin{eqnarray} \label{eq:ff}
 \bra {\gamma}\overline{u}\gamma_\alpha d\ket {\pi^-} \, &=& \, 
-\,\frac{e}{M_{\pi^+}} \, \varepsilon^{\beta *} \, F_V(q^2)
\, \varepsilon_{\alpha\beta\mu\nu} \, r^\mu \,  p^\nu \, , \nonumber \\
\bra {\gamma} \overline{u}\gamma_\alpha\gamma^5 d \ket {\pi^-}
&=&i \, \frac{e}{M_{\pi^+}}\varepsilon^{\beta *}\, F_A(q^2) \left[(r\cdot p)g_{\alpha\beta}-
p_\alpha r_\beta \right]  \nonumber \\
& &  + \, i\,e\,\varepsilon_\alpha^*\,\sqrt{2} \, F\,,
\end{eqnarray} 
where $r$ and $p$ are the pion and photon momenta, respectively, $q^2=(r-p)^2$ and $e$ is the electric
charge of the electron.
The second term in the matrix element of the axial-vector current corresponds to the pion
pole contribution (which coupling is given by the decay constant of the pion $F$) to the IB amplitude.
\par
For the full set of expressions for the differential decay rate of $\pi \rightarrow \ell \, \nu_{\ell} \gamma$
in terms of the vector and axial-vector form factors see, for instance, Ref.~\cite{Bryman:1982et}.
\par
Form factors drive the hadronization of QCD currents and embed non-perturbative aspects that we still
do not know how to evaluate from the underlying strong interaction theory. Their determination is
all-important in order to disentangle those aspects. It is reasonable to assume, as has been common lore in the 
literature on this topic, that hadronic resonance states should dominate the structure
of form factors and, accordingly, meromorphic functions with poles in the relevant resonances coupled
to the corresponding channels have been extensively proposed in order to fit hadronic data. This procedure by
itself is, however, not fully satisfactory because it does not impose known QCD constraints.
\par
On one side chiral symmetry of massless QCD drives the very low-energy region of 
form factors \cite{Weinberg:1978kz}. Hence the latter have to satisfy its constraints
in this energy domain. On the other, one can also demand that form factors in the resonance
energy region should match short-distance QCD properties. This idea was pioneered
by Ref.~\cite{Ecker:1989yg} and has been used extensively in the last years
\cite{Peris:1998nj,Knecht:2001xc,Ruiz-Femenia:2003hm,Bijnens:2003rc,Cirigliano:2004ue,otherGF}.
\par
In the following we apply these techniques in order to determine the Standard Model description
of the vector and axial-vector form factors in the radiative pion decay.
Their definition, given by Eq.~(\ref{eq:ff}), illustrates
the fact that they follow from three-point Green functions of the corresponding QCD currents. 
The proper GF in this case, namely $\braket{VVP}$ and $\braket{VAP}$, happen to be order
parameters of the spontaneous breaking of chiral symmetry hence free of perturbative contributions
in the chiral limit. This is a key aspect required by our procedure. Hence
we consider in this article their study in the chiral limit, that otherwise should
provide the dominant features.
In the following we handle the GF in order to provide a description constrained by QCD
and then we will work out the form factors. 

\subsection{Vector form factor}
The relevant GF is the $\braket{VVP}$ defined by~:
\begin{eqnarray} \label{eq:vvpgff}
\left( \Pi_{VVP} \right)_{\mu \nu}^{abc} (p,q) \, = \, && \\ &&
\! \! \! \! \! \! \! \! \! \!\! \! \! \! \!\! \! \! \! \!
\! \! \! \! \! \! \! \! \! \!\! \! \! \! \!\! \! \! \! \!
\! \! \! \! \! \! \! \! \! \!  i^2 
\int   d^4x \, d^4y \, e^{i(p \cdot x + q \cdot y)} \, 
\langle 0 | T \left\{ V_{\mu}^a(x) \, V_{\nu}^b(y) \, P^c(0) \right\} |
0 \rangle \; , \nonumber 
\end{eqnarray}
where~:
\begin{equation} \label{eq:currents1}
V_{\mu}^a(x)  =   \left( \overline{\psi} \, \gamma_{\mu} 
\frac{\lambda^a}{2} \, 
\psi \right)(x) \,   ,  \; \; 
 P^a(x) =  \left( \overline{\psi} \,
i \gamma_5  \lambda^a  \psi \right) (x)\, ,
\end{equation}
with $a=1,...,8$, octets of vector and pseudoscalar currents.
Moreover $SU(3)_V$ symmetry, parity and time reversal demand that~:
\begin{equation}
\left( \Pi_{VVP} \right)_{\mu \nu}^{abc} (p,q)  =  d^{abc}
\varepsilon_{\mu \nu \alpha \beta}  p^{\alpha}  q^{\beta}  \, 
\Pi_{VVP}(p^2,q^2,r^2)  , 
\end{equation} 
with $r_{\mu}=(p+q)_{\mu}$. 
In addition Bose symmetry requires $\Pi_{VVP}(p^2,q^2,r^2)=\Pi_{VVP}(q^2,p^2,r^2)$.
\par
The vector form factor $F_V(q^2)$, defined by Eq.~(\ref{eq:ff}), derives from 
$\Pi_{VVP}(p^2,q^2,r^2)$ in the chiral limit by~:
\begin{equation} \label{eq:fv}
 F_V(q^2) = \frac{\sqrt{2} \, M_{\pi^+}}{6 \, F B_0} \lim_{p^2,r^2\to 0} r^2 \,\Pi_{VVP}(p^2,q^2,r^2) \, ,
\end{equation}
with $B_0 = - \left\langle \overline{\psi}\psi\right\rangle_0 / F^2$.
\par
We do not know how to determine $\Pi_{VVP}(p^2,q^2,r^2)$ ($\Pi_{VVP}$ for short) from QCD in all energy domains
but our knowledge of the strong interaction theory allows us to know this function in
two specific limits as we will comment now \cite{Knecht:2001xc,Ruiz-Femenia:2003hm,Moussallam:1994xp}~: 
\begin{itemize}
 \item[i)] {\em Very low-energy region}. The GF has to satisfy the constraints of chiral symmetry
encoded in $\chi$PT. The leading ${\cal O}(p^4)$ contribution in the chiral expansion is provided by the 
Wess-Zumino anomaly \cite{Wess:1971yu} and gives~:
\begin{equation} \label{eq:chiral}
\Pi_{VVP}(p^2,q^2,r^2)  \simeq   \frac{B_0}{r^2}  \left(  
\frac{N_C}{4 \, \pi^2}  +  {\cal O}(p^2,q^2,r^2)  \right)   ,
\end{equation}
\item[ii)] {\em Asymptotic energy region}. The perturbative QCD determination of the GF, within the
Operator Product Expansion (OPE) framework gives, in the chiral limit and at ${\cal O}(\alpha_S^0)$
\footnote{Although this result is completely symmetric in the three momenta, this does not longer hold
when gluon corrections are included \cite{mateu-jamin-pich}. Hence we do not expect that this 
symmetry will be sustained when constructing the GF in the resonance energy region.}~:
\begin{eqnarray} \label{eq:cond1}
\lim_{\lambda \rightarrow \infty} \, \Pi_{VVP} ((\lambda p)^2, 
(\lambda q)^2, (\lambda p + \lambda q)^2)  = && \\
 -  \, \frac{B_0 F^2}{\lambda^4}
\, \frac{p^2 + q^2+r^2}{p^2 q^2 r^2}  +  
{\cal O}\left(\frac{1}{\lambda^6}\right) \, , \nonumber 
\end{eqnarray}
\begin{eqnarray} \label{eq:cond2}
\lim_{\lambda \rightarrow \infty} \, \Pi_{VVP} ((\lambda p)^2, (q-\lambda p)^2,
q^2)  = && \\
-\, \frac{2 B_0 F^2}{\lambda^2} \, \frac{1}{p^2 q^2} \, + \, 
{\cal O}\left(\frac{1}{\lambda^3}\right) \, , \nonumber
\end{eqnarray}
\begin{eqnarray} \label{eq:cond3}
\lim_{\lambda \rightarrow \infty} \, \Pi_{VVP} ((\lambda p)^2, q^2,
(q + \lambda p)^2)  = && \\
 i \, \frac{2}{\lambda^2} \, \frac{1}{p^2} \, 
\Pi_{VT}(q^2) \, + \, 
{\cal O}\left(\frac{1}{\lambda^3}\right) \, , \nonumber
\end{eqnarray}
where
\begin{eqnarray} \label{eq:cond4}
\delta^{ab} \, \left( \Pi_{VT} \right)_{\mu \rho \sigma}(p)  & =  & \nonumber \\ 
&& 
\! \! \! \! \! \! \! \! \! \!\! \! \! \! \!\! \! \! \! \!
\! \! \! \! \! \! \! \! \! \!\! \! \! \! \!\! \! \! \! \!
i\,
\int \, d^4x \, e^{i p\cdot x} \, \langle 0 | T \left\{ \, V_{\mu}^a(x) \, 
  T^b_{\rho\sigma}(0) \, \right\} |0 \rangle \; , \nonumber \\  \nonumber \\
\left( 
\Pi_{VT} \right)_{\mu \rho \sigma}(p) & = & \left( \, p_{\rho} \, g_{\mu \sigma}
\, - \, p_{\sigma} \, g_{\mu \rho} \, \right) \, \Pi_{VT}(p^2) \; , \nonumber
\\  \nonumber \\
\lim_{\lambda \rightarrow \infty} \, \Pi_{VT}((\lambda p)^2) & = & 
 \,i\,\frac{B_0 F^2}{\lambda^2} \, \frac{1}{p^2}  \, + \, {\cal O}
 \left( \frac{1}{\lambda^4} \right) \; , \nonumber \\
 T^a_{\rho\sigma}(x)&=& \left( \bar{q}\,\sigma_{\rho\sigma}\dfrac{\lambda^a}{2}q \right)(x)\; ,
\end{eqnarray}
where $(\Pi_{VT})_{\mu \rho \sigma}(p)$ stands for the Vector Tensor GF~\footnote{We use
$\sigma_{\mu \nu} = (i/2)\left[ \gamma_{\mu} , \gamma_{\nu} \right]$.}. 
\end{itemize}
In addition to these constraints on the $\braket{VVP}$ GF there is also a requirement
that we will enforce in any hadronic form factor of vector or axial-vector QCD currents. It is known
\cite{Floratos:1978jb} that the leading perturbative contribution, within QCD, to the spectral functions 
of both vector and axial-vector correlators is constant. Then it comes out, as a heuristic deduction,
that any of the infinite hadron contributions to the spectral functions should vanish at high transfer
of momentum. This implies, in order, that hadron form factors of those currents should behave smoothly
at high energy \cite{Rosell:2006dt}. Incidentally this feature coincides with the known
Brodsky-Lepage condition on form factors (derived within a partonic framework) \cite{Lepage:1980fj}.
Specifically the condition, in our case, reads from Eq.~(\ref{eq:fv}) as~:
\begin{equation} \label{eq:cond5}
\lim_{\begin{array}{c} p^2,r^2 \rightarrow 0 \\ q^2 \rightarrow \infty \end{array}}
r^2 \, \Pi_{VVP}(p^2,q^2,r^2) \, = \, 0 \; .
\end{equation}
Our task is to construct a function for $\braket{VVP}$ that satisfies, at least, the set of conditions
in Eqs.~(\ref{eq:chiral},\ref{eq:cond1},\ref{eq:cond2},\ref{eq:cond3},\ref{eq:cond4},\ref{eq:cond5}).
\par
To proceed we use the ideas of the large-$N_C$ limit of QCD \cite{'tHooft:1973jz} that, essentially,
tell us that GF of currents should be described, in the $N_C \rightarrow \infty$ limit, by meromorphic
functions emerging from a theory with an infinite hadron spectrum of stable states. This setting 
is difficult to handle and, in practice, one cuts the spectrum reducing it to the lightest Goldstone or
resonance mesons that obey specific hints \footnote{This method is known as Minimal Hadronic Ansatz (MHA)
\cite{Peris:1998nj}.}.
\par
In Ref.~\cite{Ruiz-Femenia:2003hm} a Lagrangian theory, including one multiplet of vector resonances only,
was designed in order to obtain an expression for $\Pi_{VVP}$ that satisfied all conditions but for the one in Eq.~(\ref{eq:cond5}). Indeed the fact that only one multiplet of vector resonances was not
enough in order to satisfy all short-distance constraints for this GF was already previously noticed \cite{Knecht:2001xc} with the use of a parametric {\em ansatz}. It is already well known 
\cite{Knecht:2001xc,Bijnens:2003rc}
that the MHA is more involved if we want that our representation of the GF satisfies both OPE and the
Brodsky-Lepage requirements. 
\par
The obvious extension is to extend our spectrum by including also a multiplet of pseudoscalar resonances
in the construction of the $\Pi_{VVP}$ function. Although it can be shown that indeed this parameterization
satisfies conditions in Eqs.~(\ref{eq:chiral},\ref{eq:cond1},\ref{eq:cond5}), it fails to meet
the OPE condition in Eq.~(\ref{eq:cond2}) \footnote{Constraints in Eqs.~(\ref{eq:cond3},\ref{eq:cond4}) are,
in this case, undetermined.}. It is not difficult to relate this problem to the fact that the $\braket{AP}$ correlator
in the chiral limit is saturated by one pion exchange \cite{mateu-jamin-pich}.
\par
It can be seen that all conditions are met if we consider in the spectrum of the $\braket{VVP}$ GF
two non-degenerated multiplets of vector resonances, together with the Goldstone pseudoscalar mesons.
Then the {\em ansatz} would read~:
\begin{eqnarray} \label{ansatz_2}
\Pi_{VVP}^{res}(p^2,q^2,r^2)   =  && \\ &&
\! \! \! \! \! \! \! \! \! \!\! \! \! \! \!\! \! \! \! \!
\! \! \! \! \! \! \! \! \! \!\! \! \! \! \!\! \! \! \! \!
 \! \! \! \!\! \! \! \! \!
  \frac{P_0+P_1+P_2+P_3}{(M_{V_1}^2 - q^2) \, (M_{V_1}^2-p^2) \,
(M_{V_2}^2 - q^2) \, (M_{V_2}^2-p^2) \, r^2}\,,   \nonumber
\end{eqnarray}
with
\begin{equation} \label{eq:pedro_ansatz} 
P_n=\sum_{k=0}^n \sum_{l=0}^k c_{n-k,k-l,l}(r^2)^{n-k}(q^2)^{k-l}(p^2)^{l} \, ,
\end{equation} 
where the coefficients are symmetric under interchange of the last two indices $c_{kml}=c_{klm}$ and
$M_{V_i}$ $i=1,2$ are the masses of two nonets of vector resonances in the 
$N_C \rightarrow \infty$ limit~\footnote{Hadron phenomenology of the pion vector form factor suggests that
the mass of the lightest vector multiplet in this limit is very well approximated by the $\rho(770)$ mass. 
Hence we will take $M_{V_1} = M_{\rho(770)}$.}.
In order to satisfy the short-distance constraints several conditions on the $c_{ijk}$ coefficients
arise. Hence the chiral symmetry behavior (\ref{eq:chiral}) gives~:
\begin{equation} \label{eq:c000}
c_{000} \, = \, B_0 \, M_{V_1}^4 M_{V_2}^4 \, \frac{N_C}{4 \pi^2} \, .
\end{equation}
The OPE condition in Eq.~(\ref{eq:cond1}) is satisfied
for~:
\begin{eqnarray} \label{eq:consvv}
c_{300} & = & c_{030} \, = \, c_{120} \, = \, c_{210} \, = \, 0 \; , 
\nonumber \\
c_{021} & = & c_{111} \, = \,- \,  B_0 \, F^2 \; .
\end{eqnarray}
Finally the Brodsky-Lepage behavior on the vector form factor,
defined by condition (\ref{eq:cond5}), 
fixes one additional parameter, namely~:
\begin{equation} \label{eq:consvv1}
 c_{020} = 0 \, .
\end{equation}
Our ansatz, with all these constraints, satisfies also the OPE conditions
in Eqs.~(\ref{eq:cond2},\ref{eq:cond3}).
\par
If we evaluate now the vector form factor from Eq.~(\ref{eq:fv}) we get~:
\begin{equation} \label{eq:fvresult}
\! \!F_V(q^2) =   \frac{M_{\pi^+}}{3  \sqrt{2}  B_0 F 
M_{V_1}^2  M_{V_2}^2} \, \frac{c_{000}  +  
c_{010} \,  q^2}{(M_{V_1}^2 - q^2)\,(M_{V_2}^2 - q^2)}  , 
\end{equation} 
and we observe that only one parameter, $c_{010}$, has not been fixed by our procedure.
The expression for the vector form factor in the radiative pion decay given by Eq.~(\ref{eq:fvresult})
is the most general one that satisfies the short-distance constraints specified above. As the
transferred momenta in the $\pi^+ \rightarrow e^+ \nu_{e} \gamma$ process is small by comparison
with the mass of the lightest vector meson resonance, $q^2 \ll M_V^2$, it is appropriate to 
perform the relevant expansion until first order in $q^2$. Using the result for $c_{000}$ given
by Eq.~(\ref{eq:c000}) it gives~:
\begin{equation}
 F_V(q^2) =  F_V(0) \, \left[ \, 1 \, + \, \lambda_V \, \frac{q^2}{M_{\pi^+}^2}
\, + \, {\cal O}(q^4) \right] \; ,
\end{equation}
where 
$\lambda_V = \Lambda_{N_C \rightarrow \infty}^V + \Lambda_{1/N_C}^V + \cdots$ admits an expansion
in $1/N_C$ and
\begin{eqnarray} \label{eq:chiex}
F_V(0) & = &  \frac{\sqrt{2} N_C M_{\pi^+}}{24 \, \pi^2 F}\; , \nonumber \\ \; \\
 \Lambda_{N_C \rightarrow \infty}^V & = & 
\frac{M_{\pi^+}^2}{M_{V_1}^2} + \frac{M_{\pi^+}^2}{M_{V_2}^2} + M_{\pi^+}^2 \frac{c_{010}}{c_{000}}
\, . \nonumber
\end{eqnarray}
We must compare our value\footnote{In the following numerical determinations we will 
use $F= 0.0924 \, \mbox{GeV}$, $M_{\pi}= 0.138 \, \mbox{GeV}$,
$M_K = 0.496 \, \mbox{GeV}$, $M_{\pi^0} = 0.135 \, \mbox{GeV}$, $M_{\pi^+} = 0.140 \, \mbox{GeV}$
and $M_{V_1} = M_{\rho(770)} = 0.775 \, \mbox{GeV}$.} for $F_V(0) \simeq 0.0271$ with the result coming from  
$\Gamma(\pi^0 \rightarrow \gamma \gamma)$ and CVC,
$F_V(0)=0.0261(9)$, and with the recent experimental fit by the PIBETA collaboration, 
$F_V(0)=0.0259(18)$ \cite{maxim}.  We recall that our result for the vector form factor
 (\ref{eq:fvresult}) arises from a large-$N_C$ procedure where 
a model of the $N_C \rightarrow \infty$ has been implemented, namely the cut in the spectrum.
At $q^2 \ll M_V^2$ this form factor has been studied up to ${\cal O}(p^6)$ in $\chi$PT 
\cite{Ametller:1993hg,Bijnens:1996wm}.
At ${\cal O}(p^4)$ the Wess-Zumino Lagrangian determines $F_V(0)$ as given in Eq.~(\ref{eq:chiex}).
Higher chiral order corrections to this result vanish in the chiral limit, accordingly their size 
is suppressed over the leading order by powers of $M_{\pi}^2/M_V^2$ or 
$M_{\pi}^2/\Lambda_{\chi}^2$ that are tiny. Indeed, using the ${\cal O}(p^6)$ odd-intrinsic parity
Lagrangian ${\cal L}_6^W$ worked out in Ref.~\cite{Bijnens:2001bb}, this modification to $F_V(0)$ is
proportional to a low-energy constant as $M_{\pi^+}^2 \, C_7^W$, that also contributes to the 
$\pi^0 \rightarrow \gamma \gamma$ decay. From the latter one obtains \cite{Strandberg:2003zf} 
$C_7^{W} \simeq (0.013 \pm 1.17) \times 10^{-3} \, \mbox{GeV}^{-2}$, {\it i.e.} compatible with zero.
\par
The slope $\lambda_V$ arises at ${\cal O}(p^6)$ 
with the usual two features~: The local operator $O_{22}^W$ in ${\cal L}_6^W$ provides the
$N_C \rightarrow \infty$ contribution~:
\begin{equation} \label{eq:c22lambda}
 \Lambda_{N_C \rightarrow \infty}^{\chi PT} \, = \, \frac{64 \, \pi^2}{N_C} \, 
M_{\pi^+}^2 \,C_{22}^{Wr}(\mu)\, ,
\end{equation}
that is ${\cal O}(1)$ in the large-$N_C$ expansion,
and a one-loop calculation that provides the chiral logs corresponding to 
the next-to-leading order~\cite{Ametller:1993hg}~:
\begin{equation}
\Lambda_{1/N_C}^{\chi PT} = \, - \,\frac{M_{\pi^+}^2}{48 \, \pi^2 F^2} \left[ 1 + \log \left( \frac{M_{\pi}^2}{\mu^2}
\right) \right] \, . 
\end{equation}
There is another process directly related with the $\braket{VVP}$ GF, 
namely $\pi \rightarrow \gamma \gamma^*$; hence it should be related with the radiative pion decay.
Indeed within the assumptions that carried us to $F_V(q^2)$ in Eq.~(\ref{eq:fvresult}), the momenta
structure for the $\pi \rightarrow \gamma \gamma^*$ decay should be the same, though with different
normalization. In consequence the $\Lambda_{N_C \rightarrow \infty}^V$ slope in Eq.~(\ref{eq:chiex}) is
the same for both processes.
\par
The $\pi^0 \rightarrow \gamma e^+ e^-$ amplitude can be expressed by~:
\begin{equation}
 {\cal M}_{\pi \rightarrow \gamma \gamma^*} \, = \, 
{\cal M}_{\pi \rightarrow \gamma \gamma}  \, \left[ \, 1 \, + \, \lambda_{\gamma} \,
\frac{q^2}{M_{\pi^0}^2} \, + ... \right] \, ,
\end{equation}
where $q^2 = (p_{e^+} + p_{e^-})^2$. The slope arises at ${\cal O}(p^6)$ in $\chi$PT
and it is \cite{Bijnens:1989jb} \footnote{Notice that one-loop ${\cal O}(p^6)$ $\chi$PT contributions, 
encoded in $\Lambda_{1/N_C}^{\chi PT}$,
coincide in $\pi \rightarrow e \, \nu_{e} \gamma$ and $\pi \rightarrow \gamma \gamma^*$
in the $SU(3)_V$ limit, {\it i.e.} for $M_K = M_{\pi}$.}~:
\begin{equation}
 \lambda_{\gamma} \, = \, \frac{64  \pi^2}{N_C}   M_{\pi^0}^2 \,C_{22}^{Wr}(\mu)\, 
- \, \frac{M_{\pi^0}^2}{96 \pi^2  F^2} \left[ 2  +  \log 
\left(\frac{M_{\pi}^2 M_K^2}{\mu^4} \right) \right].
\end{equation}
Fortunately it has been measured rather accurately \cite{Yao:2006px}, 
$\lambda_{\gamma} = 0.032 \pm 0.004$ and then we can input this measure to determine
the LEC $C_{22}^W(\mu)$ obtaining~:
\begin{equation}
 C_{22}^{Wr}(M_{\rho}) \, = \, 7.0 ^{+1.0} _{-1.5} \times 10^{-3} \, \mbox{GeV}^{-2} \, , 
\end{equation}
where the error includes also the incertitude of the renormalization point $\mu$ between $M_{\rho}$
and $1 \, \mbox{GeV}$. Coming back to the slope of $F_V(q^2)$ we get~:
\begin{equation}
\lambda_V \, = \, 0.041^{+0.004}_{-0.007} \, ,
\end{equation}
that compares well with the recent PIBETA measurement $\lambda_V = 0.070 \pm 0.058$
\cite{maxim}.
\par 
By comparing now $\Lambda_{N_C \rightarrow \infty}^V$ in Eq.~(\ref{eq:chiex}) and 
$\Lambda_{N_C \rightarrow \infty}^{\chi PT}$ in Eq.~(\ref{eq:c22lambda}) we can provide a
determination for the undetermined parameter in the GF and then give a full prescription for
the $F_V(q^2)$ form factor in Eq.~(\ref{eq:fvresult}). For the mass of the 
first multiplet of vector resonances we take $M_{\rho(770)}$ and for the second 
$M_{\rho(1450)} = 1.459 \, \mbox{GeV}$~:
\begin{equation}
 \frac{c_{010}}{c_{000}} \, = \, (-0.7 \pm 0.3) \, \mbox{GeV}^{-2} \, .
\end{equation}
Notice that the size of this parameter is of the same order that the other two terms in 
$\Lambda_{N_C \rightarrow \infty}^V$. With this result we end the construction of the
vector form factor in radiative pion decays in the large-$N_C$ limit given by
Eq.~(\ref{eq:fvresult}).
\par
It is interesting to compare our results with those in 
Refs.~\cite{Ruiz-Femenia:2003hm,Ruiz-Femenia:2003gw}.
As commented above the construction of the $\braket{VVP}$ GF in those references was
carried out using only one multiplet of vector resonances, hence the vector form factor
in radiative pion decay did not satisfy the constraint in Eq.~(\ref{eq:cond5}). With this 
setting we were able to give a full prediction for the leading contribution to the
slope $\lambda_V$, namely,
\begin{equation}
 \Lambda_{N_C \rightarrow \infty}^{1R} \, = \, \frac{M_{\pi^+}^2}{M_V^2} \, 
\left[ 1 - \frac{4 \pi^2 F^2}{N_C M_V^2} \right] \, .
\end{equation}
Using $M_V = M_{\rho}$ we got $\Lambda_{N_C \rightarrow \infty}^{1R} \simeq 0.027$
to be compared with $\Lambda_{N_C \rightarrow \infty}^V = 0.028 \pm 0.006$ from our analysis
above.
\par
The study of the $\braket{VVP}$ GF along the lines outlined in this section can also be 
carried out within a resonance Lagrange theory instead of a parametric representation as
given by Eq.~(\ref{ansatz_2}). We collect this procedure in Appendix~\ref{app:a}.

\subsection{Axial-vector form factor}
We now come back to the axial-vector form factor defined by Eq.~(\ref{eq:ff}). In order to
determine the $F_A(q^2)$ form factor we follow an analogous procedure to the one outlined before
for the vector form factor. The relevant GF is, in this case, the $\braket{VAP}$ defined by~:
 \begin{eqnarray}
\left( \Pi_{VAP} \right)_{\mu \nu}^{abc} (p,q) \, = \, && \\ &&
\! \! \! \! \! \! \! \! \! \!\! \! \! \! \!\! \! \! \! \!
\! \! \! \! \! \! \! \! \! \!\! \! \! \! \!\! \! \! \! \!
\! \! \! \! \! \! \! \! \! \!  i^2 
\int   d^4x \, d^4y \, e^{i(p \cdot x + q \cdot y)} \, 
\langle 0 | T \left\{ V_{\mu}^a(x) \, A_{\nu}^b(y) \, P^c(0) \right\} |
0 \rangle \; , \nonumber 
\end{eqnarray}
where~:
\begin{equation}
A_{\mu}^a(x)  =   \left( \overline{\psi}  \, \gamma_{\mu} \gamma_5
\frac{\lambda^a}{2} \, 
\psi \right)(x) \, ,
\end{equation}
and the other currents have been defined in Eq.~(\ref{eq:currents1}). The structure
of this GF is slightly more complicated than $\braket{VVP}$ as it involves two scalar
functions~:
\begin{eqnarray}
 \left( \Pi_{VAP} \right)_{\mu \nu}^{abc} (p,q) & = & 2
f^{abc} \left\{ B_0 F^2 \left[\frac{(p+2q)_{\mu} q_{\nu}}{q^2 r^2}-\frac{g_{\mu \nu}}{r^2} \right] \right. 
\nonumber \\ 
&& \; \; \; \; \; \; \; \; \; \left. +P_{\mu \nu}(p,q) \, {\cal F}(p^2,q^2,r^2) \right. \\ 
&&  \; \; \; \; \; \;\; \;  \; +Q_{\mu \nu}(p,q) \, {\cal G}(p^2,q^2,r^2) \Big\}\, , \nonumber
\end{eqnarray}
with $r_{\mu} = (p+q)_{\mu}$ and
\begin{eqnarray}
 P_{\mu \nu}(p,q) & = & q_{\mu} p_{\nu} - ( p \cdot q)g_{\mu \nu} \; , \\
Q_{\mu \nu}(p,q) & = & p^2 q_{\mu} q_{\nu} + q^2 p_{\mu} p_{\nu}-(p \cdot q)p_{\mu} q_{\nu} -
p^2 q^2 g_{\mu \nu} \; . \nonumber
\end{eqnarray}
Then the axial-vector form factor is obtained through~:
\begin{equation}
 F_A(q^2) \, = \, \frac{\sqrt{2} M_{\pi^+}}{B_0 \, F} \lim_{p^2,r^2 \rightarrow 0}
r^2 \, {\cal F}(p^2,q^2,r^2) \; .
\end{equation}
A detailed study of this GF along the line we have performed above for the $\braket{VVP}$ function
was performed in Ref.~\cite{Cirigliano:2004ue}. One of the conclusions achieved was that the 
inclusion of one multiplet of vector, axial-vector and pseudoscalar resonances (together with
the pseudoscalar mesons) was enough to satisfy the matching to the OPE expansion of the 
$\braket{VAP}$ GF at leading order. Moreover the analogous to the Brodsky-Lepage condition
(\ref{eq:cond5}), in this case, was also satisfied, {\it i.e.} the resulting axial-vector form factor
$F_A(q^2)$ behaves smoothly at high $q^2$. We refer the reader to that reference for details. 
Hence we obtain, for $N_C \rightarrow \infty$ with a cut spectrum~:
\begin{equation} \label{eq:faresult}
F_A(q^2)=\frac{\sqrt{2} F M_{\pi^+} }{M_A^2-q^2} \left( \frac{M_A^2}{M_V^2} - 1 \right) \,,
\end{equation}
where $M_A$ is the mass of the lightest axial-vector multiplet of resonances in the $N_C \rightarrow \infty$
limit.

\begin{table*}
\caption{Comparison of theoretical and experimental determinations for the
low-energy expansion of vector and axial-vector form factors. The PIBETA determination assumes that the axial-vector form factor is constant, {\it i.e.} it does not consider a slope. }
\label{tab:tablahadronic}
\renewcommand{\arraystretch}{1.3}\setlength{\LTcapwidth}{\textwidth}
\begin{center}
\begin{tabular}[tbh]{|c|c|c|c|c|}
\hline 
&Experiment \cite{Pokanic06}& $SU(2)$ Ref.~\cite{Bijnens:1996wm}& $SU(3)$ Ref.~\cite{Geng:2003mt}& Our work\\
\hline 
$F_V(0)$   & 0.0258(18)      & 0.0271 & 0.0272  & 0.0271  \\\hline 
$\lambda_V$& 0.070(58)       & 0.044  & 0.045   & 0.041   \\\hline 
$F_A(0)$   & 0.0121(18)      & 0.0091 & 0.0112  & exp. input \\\hline 
$\lambda_A$& not measured    & 0.0034 & $\sim\,$0 & 0.0197(19)\\
\hline 
\end{tabular}
\end{center}
\end{table*}
At $q^2 \ll M_A^2$ we may resort again to $\chi$PT \cite{Bijnens:1996wm,Geng:2003mt} with the 
expansion~:
\begin{equation}
 F_A(q^2) \, = \, F_A(0) \, \left[ 1 + \lambda_A \, \frac{q^2}{M_{\pi^+}^2} + ... \right] \, .
\end{equation}
Both terms, $F_A(0)$ and slope, satisfy an expansion in $1/N_C$, for instance 
$\lambda_A = \Lambda_{N_C \rightarrow \infty}^A + \Lambda_{1/N_C}^A + ...$.
From our result above we get~:
\begin{eqnarray} \label{eq:anc}
 F_A(0) & = & \sqrt{2} F M_{\pi^+} \left( \frac{1}{M_V^2} - \frac{1}{M_A^2} \right) \, , \nonumber \\
\Lambda_{N_C \rightarrow \infty}^A & = & \frac{M_{\pi^+}^2}{M_A^2} \, .
\end{eqnarray}
$F_A(q^2)$ arises first at ${\cal O}(p^4)$ with a constant local contribution from the $\chi$PT Lagrangian,
namely~:
\begin{equation} \label{eq:fa4q2}
 F_A^{(4)}(q^2) \, = \, 4 \sqrt{2} \, \frac{M_{\pi^+}}{F} \left( L_9^r + L_{10}^r \right).
\end{equation}
The next corrections appear at ${\cal O}(p^6)$ in the chiral expansion \cite{Geng:2003mt}. 
One of them results from local operators of the ${\cal O}(p^6)$ chiral Lagrangian that, in the chiral limit, only 
contribute to $\lambda_A$~: 
\begin{equation} \label{eq:cila}
 \lambda_A^{(6)} |_{N_C \rightarrow \infty} \, = \, 
\frac{M_{\pi^+}^2}{L_9^r+L_{10}^r} \left[ C_{78}^r - 2 C_{87}^r + 
\frac{1}{2} C_{88}^r \right] \, .
\end{equation}
There is also a subleading term, in the large-$N_C$ expansion, that comes from one-loop
diagrams involving the ${\cal O}(p^4)$ chiral Lagrangian. However it only affects $F_A(0)$ and it is zero in the
chiral limit. 
The third correction is sub-subleading and results from two-loop diagrams evaluated with the ${\cal O}(p^2)$ chiral Lagrangian. The latter contributes both to $F_A(0)$ and $\lambda_A$.
All local additions, $F_A^{(4)}(q^2)$ and $\lambda_A^{(6)} |_{N_C \rightarrow \infty}$, 
correspond to our result in Eq.~(\ref{eq:anc}), {\it i.e.} $N_C \rightarrow \infty$, when LECs are saturated 
by resonance contributions \cite{Cirigliano:2004ue}. 
Though the full ${\cal O}(p^6)$ chiral result is rather cumbersome, the authors of Ref.~\cite{Geng:2003mt}
have provided a numerical expression for the renormalization scale $\mu= M_{\rho}$. The conclusion is that, in the chiral limit, subleading contributions to the slope are negligible. Notwithstanding it is relevant to emphasize that both $\chi$PT results of Refs.~\cite{Bijnens:1996wm,Geng:2003mt} use models to evaluate the resonance contributions to the ${\cal O}(p^6)$ local terms and, accordingly, their final conclusion is tamed by this estimate.
\par
We turn now to give our numerical results. Contrarily to what happens in the vector case, where 
the lightest vector resonance mass in the $N_C \rightarrow \infty$ limit is well approximated by
the $\rho(770)$ mass, the axial-vector mass in that limit ($M_A$) differs appreciably from the
lightest multiplet of these resonances, namely $a_1(1260)$. The result $M_A = \sqrt{2} M_V$ was 
obtained in Ref.~\cite{Ecker:1989yg} by imposing several short-distance constraints on the couplings
of the resonance Lagrangian. Lately \cite{Rosell:2006dt} it has been noticed that the inclusion 
of NLO effects in the large-$N_C$ expansion points out to $M_A \leq \sqrt{2} M_V$. These results
are rather different from the mass of the lightest axial-vector meson determined experimentally 
$M_A \simeq 1.230 \, \mbox{GeV} \simeq M_{a(1260)}$~\cite{GomezDumm:2003ku} but it is important to
remind that this resonance is rather wide.
\par
Our strategy is the following~: we will use the experimental value of $F_A(0)$ as given in
Tab.~\ref{tab:tablahadronic} to determine $M_A$ through Eq.~(\ref{eq:anc}); then we provide 
a prediction for $\lambda_A$. We find~\footnote{It is important to notice that the value of 
$F_A(0)$ measured by the PIBETA experiment assumes no slope for the axial-vector form factor. We
should repeat this exercise when $\lambda_A$ is included.}~:
\begin{eqnarray}\label{eq:mala}
 M_A &= & 998 \, (49) \, \mbox{MeV} \, , \nonumber \\ 
\lambda_A & = & 0.0197 \, (19) \, ,
\end{eqnarray}
where the error stems only from the experimental uncertainty in $F_A(0)$. Notice that this
result satisfies $M_A \leq \sqrt{2} M_V \simeq 1096 \, \mbox{MeV}$.

\subsection{Theory versus Experiment} \label{subsect:error}
We are now ready to compare our results with other theoretical settings and experimental determinations.
In Tab.~\ref{tab:tablahadronic} we compare our outcome for the low-energy expansion of the form factors
with the one provided by ${\cal O}(p^6)$ $\chi$PT and the recent PIBETA published values.
\par
As $F_V(0)$ is ruled by the Wess-Zumino anomaly all the theoretical results agree for this parameter.
Leading corrections to this value are driven by the pion mass and as a result happen to be tiny 
\cite{Ruiz-Femenia:2003hm}. This is also reflected
in the excellent comparison with the experimental determination. The agreement is also good for the slope
of the vector form factor, considering the large error of the experimental value.
\par
The axial-vector form factor does not arise a similar consensus. As indicated above
$\chi$PT can only predict reliably all loop contributions (up to ${\cal O}(p^4)$ in the even-intrinsic parity and ${\cal O}(p^6)$ in the odd-intrinsic parity sectors) while higher order loops involve the couplings of local operators. Moreover tree-level
${\cal O}(p^4)$ (\ref{eq:fa4q2}) and ${\cal O}(p^6)$ (\ref{eq:cila}) terms can only be determined in different models for resonance saturation contributions. 
The excellent agreement between the $\chi$PT results and the experimental determination of $F_A(0)$ is, indeed, not a major issue as the axial-vector form factor in radiative pion
decay is the main phenomenological source\footnote{$L_9^r$ is rather well
determined from the phenomenology (squared charge radius of the pion) and its numerical value agrees nicely with resonance saturation.} to fix the value of $L_{10}^r$ . It happens that $F_A^{(4)}(0)$ arises from a strong cancellation between the $L_9^r$ and $L_{10}^r$ LECs and, in consequence, it is very sensitive to the chosen value for $L_{10}^r$.  In terms of resonance saturation this sensitivity
moves to the value of the axial-vector mass $M_A$ input in the numerical determination. The value of 
$L_{10}^r \simeq -5.5 \times 10^{-3}$, used by Ref.~\cite{Geng:2003mt}, arises for $M_A \simeq 1 \, \mbox{GeV}$.
\par
Our model of large-$N_C$ gives the leading result for the axial-vector form factor parameters and there 
are leading Goldstone-mass driven contributions that we have not considered. In the $\chi$PT framework these ${\cal O}(p^6)$ corrections arise from the LECs and, a priori, it is difficult to estimate their contribution
due to our lack of reliable knowledge on those low-energy couplings. However it has been pointed out
\cite{Geng:2003mt} that the role of LECs is unimportant in the
${\cal O}(p^6)$ corrections. As the subleading loop contributions are also tiny, it is concluded that $\lambda_A$ is not sizeable and $F_A(0)$ is ruled by the leading ${\cal O}(p^4)$ contribution by far.
\par
However using as input the experimental value of $F_A(0)$ we find a large value for $\lambda_A$. As
subleading $1/N_C$ loop contributions seem to be tiny our leading result shows a clear discrepancy
with the estimates of tree-level contributions performed in the chiral framework 
\cite{Bijnens:1996wm,Geng:2003mt}. It would be very much interesting to have an experimental
determination of $\lambda_A$ in order to disentangle the different resonance models.

\section{Beyond SM~: Tensor form factor} \label{sec:beyond}
As pointed out in the Introduction, the history of the radiative decay of the pion accumulates 
a few clashes between Theory and Experiment. It seems though that, after the latest analysis by the 
PIBETA collaboration, the landscape has very much soothed. However it has become customary to
investigate possible contributions beyond the Standard Model in order to appease alleged discrepancies.
Between the latter the possible role played by a tensor form factor has thoroughly been studied
\cite{Poblaguev:1990tv,Belyaev:1991gs,Herczeg:1994ur,Poblaguev:2003ib,Chizhov:2004tu}. 
\par
The new short-distance interaction can be written in terms of quark and lepton currents 
and it reads~:
\begin{equation} \label{eq:snew_hort}
\mathcal{L}_T=\dfrac{G_F}{2\sqrt{2}}\,V_{ud} \, F_T\,\left[ \, \bar{q}\,\sigma_{\mu\nu}\,(1-\gamma_5)\,q \, \right]
\,\left[ \, \bar{\ell}\,\sigma^{\mu\nu}\,(1-\gamma_5)\,\nu_{\ell} \, \right]\,, 
\end{equation}
where $F_T$ is an adimensional parameter measuring the strength of the new interaction. As the product $\sigma^{\mu\nu}\,\gamma^5$ is not an independent Dirac matrix (due to the identity
$\sigma^{\mu\nu}\,\gamma^5\,=\,-\,\frac{i}{2}\varepsilon^{\mu\nu\alpha\beta}\,\sigma_{\alpha\beta}$) we can write (\ref{eq:snew_hort}) as~:
\begin{equation}
\mathcal{L}_T=-\dfrac{G_F}{\sqrt{2}}\,V_{ud} \, F_T\,\left[ \,  \bar{q}\,\sigma_{\mu\nu}\,\gamma_5 \,q \, \right]
\,\left[ \, \bar{\ell}\,\sigma^{\mu\nu}\,(1-\gamma_5)\,\nu_{\ell}\, \right]\,.
\end{equation}
In the Standard Model the later structure, a tensor-like quark-lepton interaction, arises from loop
corrections to the tree-level amplitudes and gives a tiny value for $F_T \sim 10^{-8}$ \cite{Belyaev:1991gs}.
More sizeable contributions could come from New Physics models. Leptoquark exchanges, for instance, could 
give $F_T \sim 10^{-3}$ \cite{Herczeg:1994ur}, while SUSY contributions provide $F_T \sim 10^{-4}-10^{-5}$
\cite{Belyaev:1991gs} for light supersymmetric partners.
\par
The hadronization of the tensor current, at very low transfer of momenta, is driven by the constant $f_T$ 
defined by~:
\begin{equation} \label{eq:tensorhadron}
\left\langle \gamma \left| \bar{u}\,\sigma_{\mu\nu}\,\gamma_5\, d\right|\pi^-\right\rangle =
\,-\,\dfrac{e}{2}\,f_T\,(p_\mu\epsilon_\nu -p_\nu\epsilon_\mu) \, ,
\end{equation}
where $p$ is the photon momentum \footnote{There is in fact another Lorentz structure contributing to this
matrix element but it carries higher orders in momenta. If the latter is included $f_T$ acquires a 
dependence in the squared of the transferred momenta, {\it i.e.} $f_T(q^2)$. See Appendix~\ref{app:b} for a detailed evaluation of both form factors.}.
The determination of $f_T$ involves QCD in its non-perturbative regime and, consequently, is a 
non-trivial task. We will come back to this issue in the next Subsection.
\par
It is possible to obtain the product ${\cal T} = F_T \, f_T$ from the analyses of different processes. Hence
from some previous discrepancy in the  $\pi \rightarrow e \, \nu_e \gamma$ process it is found that
${\cal T}_{\pi} = -(5.6 \pm 1.7) \times 10^{-3}$ \cite{Poblaguev:1990tv}, while from the introduction of a 
Gamow-Teller term in the amplitude of nuclear $\beta$-decay \cite{Quin:1993vh} gives 
${\cal T}_N = (1.8 \pm 1.7) \times 10^{-3}$.

\subsection{$\braket{VT}$ Green function~: The tensor form factor}
If we want to extract information on the value of $F_T$ from experimental data, we need 
a reliable QCD-based determination of the hadronic tensor form factor. 
Using LSZ and at leading order in the pion mass we can express the matrix element (\ref{eq:tensorhadron})
 as follows \footnote{See Appendix~\ref{app:c} for a derivation of this expression.}~:
\begin{eqnarray} \label{eq:LSZtwice}
\left\langle \gamma \left| \bar{u}\,\sigma_{\mu\nu}\,\gamma_5\, d\right|\pi^-\right\rangle &=&\dfrac{i}{\sqrt{2}\,F}\left\langle \gamma \left|\bar{u}\sigma_{\mu\nu}u+\bar{d}\sigma_{\mu\nu}d\right|0\right\rangle=\nonumber\\
&-&i\,\dfrac{\sqrt{2}\, e}{3 \, F} \, \Pi_{VT}(0)(p_\mu\epsilon_\nu-p_\nu\epsilon_\mu),
\end{eqnarray}
where in the last step we have used again the LSZ reduction formula applied to the $\braket{VT}$ 
correlator defined in Eq.~(\ref{eq:cond4}). Then we have~:
\begin{equation} \label{eq:ftVT}
 f_T \, = \, i \, \frac{2 \, \sqrt{2}}{3 \, F} \, \Pi_{VT}(0) \, .
\end{equation}
To determine the $\braket{VT}$ correlator
it suffices to employ Eq.~(\ref{eq:cond3}) with our ansatz in Eq.~(\ref{ansatz_2}). It gives~:
\begin{equation}\label{eq:VT2mult}
\Pi_{VT}(q^2) \, = \, \frac{i}{2} \, \frac{c_{110} \,+ \, c_{200} \, +
\,  2 \, c_{111} \, q^2}{(M_{V_1}^2-q^2) \, (M_{V_2}^2 - q^2)} \; ,
\end{equation}
that satisfies the proper high-energy behaviour ruled by Eq.~(\ref{eq:cond3}). Unfortunately $\Pi_{VT}(0)$ is not fully specified by short-distance constraints within our approach as we have only fixed the $c_{111}$ 
parameter.
\par
In order to provide an estimate for $f_T$ we consider this correlator including one 
multiplet of vector resonances only. To proceed
we can use the results for the $\braket{VVP}$ GF as in Ref.~\cite{Ruiz-Femenia:2003hm} 
(using Eq.~(\ref{eq:cond3}) for instance) or, 
equivalently, from the $\braket{VAP}$ GF~\cite{Knecht:2001xc,Cirigliano:2004ue}.
Both procedures yield the same result, namely~:
\begin{equation} \label{eq:GFRCT}
\Pi_ {VT}(q^2)= -\, i\, \dfrac{B_0 \, F^2}{M_V^2-q^2}\,,
\end{equation} 
that matches the OPE result \cite{Craigie:1981jx}~:
\begin{equation} \label{eq:GFOPE}
 \lim_{\lambda\to \infty}\Pi_{VT}(\lambda^2 q^2)\, =\, i\, \dfrac{B_0 \, F^2 }{\lambda^2 \, q^2}\,+\,\mathcal{O}\left(\dfrac{1}{\lambda^4} \right) \,,
\end{equation} 
when the large momentum limit is taken. 
Notice that the result (\ref{eq:GFRCT}) can be recovered from (\ref{eq:VT2mult}) by taking the limit $M_{V_2}\to\infty$, demanding $c_{111}=0$ and identifying the rest of the constants. 
Using Eq.~(\ref{eq:GFRCT}) in Eq.~(\ref{eq:ftVT}) yields~:
\begin{equation} \label{eq:tensorchiral}
f_T \, = \, \dfrac{2\sqrt{2}\, B_0 \, F}{3 \, M_V^2}\, .
\end{equation} 
An educated guess can be obtained by writing $B_0 \, F \,= -\braket{\bar \psi \psi}_0/F $ and the estimate
$\braket{\bar \psi \psi}_0(1 \, \mbox{GeV}) = - \left(242\pm 15 \, \mbox{MeV}\right)^3$ \cite{Jamin:2002ev}.
We obtain $f_T = 0.24 \pm 0.04$.
\par
Another parameter of interest is the susceptibility of the quark condensate $\chi_z$ defined 
by the vacuum expectation value of the tensor current in the presence of an external source $Z_{\mu \nu }$ \cite{Ioffe:1983ju,He:1996wy}~:
\begin{equation}
 \left\langle \, 0 \, | \, \bar{\psi} \, \sigma_{\mu \nu} \, \psi \, | \, 0 \, \right\rangle_{Z} \, = \, 
g_{\psi} \, \chi_z \, \braket{\bar \psi \psi}_0 \, Z_{\mu \nu}\,.
\end{equation}
In our case we consider the magnetic susceptibility $\chi$ given by an external electromagnetic field as~:
\begin{equation}
 \left\langle \, \gamma \, | \, \bar u \sigma_{\mu \nu} u \, + \, \bar d \sigma_{\mu \nu} d \, | \, 0 
\, \right\rangle \, = \, - i e  \left(e_u + e_d \right) \, \chi \, \braket{\bar \psi \psi}_0 \, F_{\mu \nu} \, 
\end{equation}
with $e_u = 2/3$ and $e_d = -1/3$. Using the first equality of Eq.~(\ref{eq:LSZtwice}) we get~:
\begin{equation}
 f_T \, = \, - \, \frac{\sqrt{2}}{3} \, \chi \, B_0 F \, ,
\end{equation}
and comparing with Eq.~(\ref{eq:tensorchiral}) we obtain~:
\begin{equation}
 \chi \, = \, - \, \frac{2}{M_V^2} \, \simeq \, -3.3 \, \mbox{GeV}^{-2} \, .
\end{equation}
There are several determinations of the magnetic susceptibility that provide a range that runs from 
$\chi = -(8.16\pm0.41) \, \mbox{GeV}^{-2}$ \cite{Ioffe:1983ju} up to 
$\chi \simeq - 2.7  \, \mbox{GeV}^{-2}$ \cite{Ball:2002ps}.

\subsection{Lattice data and sum rules}
The last years have witnessed and increasing attention to the determination of 
matrix elements of tensor quark currents. For instance, together with the QCD sum rules technique
\cite{Ball:2002ps,Bakulev:1999gf}, lattice has also performed evaluations of amplitudes involving the
tensor current and a vector resonance \cite{Braun:2003jg,Becirevic:2003pn}~:
\begin{eqnarray} \label{eq:fvperp}
\left\langle 0\left| \, V_{\mu}^a \, \right|\rho^b(p,\lambda)\right\rangle &=& -\frac{1}{\sqrt{2}} \, \delta^{ab}
\, M_V \, f_V \, \epsilon_\mu^\lambda,\\
\left\langle 0\left| \,    T_{\mu \nu}^a \, \right|\rho^b(p,\lambda)\right\rangle &=& -
\frac{i}{\sqrt{2}} \, \delta^{ab}\,  f^\perp_V (\mu) \,\left( \epsilon_\mu^\lambda \, p_\nu-\epsilon_\nu^\lambda \, p_\mu\right), \nonumber
\end{eqnarray}
where $\rho^b(p,\lambda)$ is a vector resonance with momentum $p$, helicity $\lambda$ and polarization vector $\epsilon_\mu^\lambda$. The vector and tensor quark currents have been defined in 
Eqs.~(\ref{eq:currents1},\ref{eq:cond4}). The scale dependence of the $f^{\perp}_V(\mu)$ in 
Eq.~(\ref{eq:fvperp}) reflects the fact that the tensor current has a non-vanishing anomalous dimension.
\par
Within the large-$N_C$ framework it can be shown \cite{Craigie:1981jx,Oscar-Vicent} that if we consider a single multiplet of vector mesons we get~:
\begin{equation}
\Pi_{VT}(q^2)=\,-\,\frac{i}{2} \,\dfrac{f_V\,f_V^\perp\,M_V}{M_V^2-q^2}\,.
\end{equation} 
Comparing with Eq.~(\ref{eq:GFRCT}) we obtain the relation~:
\begin{equation} \label{eq:product}
 f_V \, f_V^{\perp} \, = \, \frac{2 \, B_0 \, F^2}{M_V} \, .
\end{equation}
The $f_V$ coupling can be obtained from the measured $\Gamma(\rho^0 \rightarrow e^+ e^-)$ \cite{Yao:2006px}.
We obtain $f_V \simeq 221 \, \mbox{MeV}$ with an expected tiny error \footnote{The vector coupling can also
be determined from short-distance analyses within resonance theory \cite{Cirigliano:2004ue}, giving
$
 f_V^2  =  2  \frac{F^2 M_A^2}{M_A^2-M_V^2},
$
that translates into $f_V = 207 \,(15) \, \mbox{MeV}$ for $M_A = 998 \,(49) \, \mbox{MeV}$ (\ref{eq:mala}), in excellent agreement with the quoted phenomenological result.}.
Then from Eq.~(\ref{eq:product}) and using the value of the quark condensate quoted above we get~:
\begin{equation} \label{eq:fvperpnu}
 f_V^{\perp}\left(1 \, \mbox{GeV}\right) \, = \, 165 \pm 31 \, \mbox{MeV} \, ,
\end{equation}
where the error collects only the uncertainty in the value of $\braket{\bar \psi \psi}_0$.
Our result is in excellent agreement with those coming from QCD sum rules~:
$f_\rho^{\perp} = 160 (10) \, \mbox{MeV}$ \cite{Ball:2002ps} and 
$f_\rho^{\perp} = 157 (5) \, \mbox{MeV}$ \cite{Bakulev:1999gf}.
\par
Lattice evaluations determine the ratio with the vector coupling. From our results we get
\begin{equation} \label{eq:matchOPE}
 \frac{f_V^{\perp}}{f_V} \left( 1 \, \mbox{GeV} \right) \, = \, 0.75 \pm 0.14 \, .
\end{equation}
to be compared with the quenched value \cite{Braun:2003jg,Becirevic:2003pn}, run down to 
$\mu = 1 \, \mbox{GeV}$~:
\begin{equation} 
 \frac{f_\rho^{\perp}}{f_\rho} \left( 1 \, \mbox{GeV} \right) \, = \, 0.74 \pm 0.03 \, .
\end{equation}
Finally from the later result and the phenomenological value of $f_V$ lattice provides the
determination~:
\begin{equation}
 f_\rho^{\perp} \left( 1 \, \mbox{GeV} \right) \, = \, 164 \pm 7 \, \mbox{MeV} \, ,
\end{equation}
to compare with our figure in Eq.~(\ref{eq:fvperpnu}).

\begin{table*}
\caption{Comparison of the theoretical predictions and the experimental data for ${\cal R}_Q = 10^8 \,R_Q$ for constant form-factors and different predictions of the $q^2$ dependence.}\label{tab:tabladatos}
\renewcommand{\arraystretch}{1.3}\setlength{\LTcapwidth}{\textwidth}
\begin{center}
\begin{tabular}[tbh]{|c|c|c|c|c|c|c|c|}
\hline $E_{e^+}^{\mathrm{min}}$(MeV) & $E_{\gamma}^{\mathrm{min}}$(MeV) &
$\theta_{e\gamma}^{\mathrm{min}}$ & ${\cal R}_{\mathrm{exp}}$ \cite{maxim} 
&${\cal R}_{\mathrm{th}}$ (without slopes)&${\cal R}_{\mathrm{th}}$ (with slopes)&${\cal R}_{\mathrm{th}}$ SU(2) \cite{Bijnens:1996wm}&${\cal R}_{\mathrm{th}}$ SU(3) \cite{Geng:2003mt}    \\
\hline 50 & 50 & --             & 2.614(21) & 2.78(38)  & 2.81(38)  & 2.46(35)  & 2.72(38) \\
\hline 10 & 50 & $40\,{}^\circ$ & 14.46(22) & 14.81(54) & 15.08(58) & 14.73(53) & 15.00(57)\\
\hline 50 & 10 & $40\,{}^\circ$ & 37.69(46) & 38.08(98)  & 38.41(103)  & 37.51(94) & 38.17(103)\\
\hline 
\end{tabular}
\end{center}
\end{table*}

\section{Analysis of the photon spectrum in the radiative pion decay}
The PIBETA experiment has thoroughly measured the photon spectrum in the radiative decay
of the pion \cite{Frlez:2003pe}. Though the results of that reference seemed to confirm
a serious discrepancy with theoretical determinations, an ensuing analysis of more data and
the refinement of systematic errors
\cite{Pokanic06,maxim} has brought a close agreement between Theory and Experiment. 
\par
The experimental available data amounts to the branching ratio of the radiative pion decay
integrated in different subregions ($Q$) of the final state phase space~:
\begin{equation}
R_Q=\dfrac{1}{\Gamma_{\pi\to e\,\nu}}\int_Q dQ_3 \sum_{\lambda} \left| \mathcal{M}(E_e,E_\nu)\right|^2\,,
\end{equation}  
where the sum runs over the polarizations of the final particles. The three regions and the experimental results are shown in Tab.~\ref{tab:tabladatos}. 
\par
We test the predictions ruled by our determination for the hadronic form factors with the experimental data, ignoring first a possible tensor interaction, and compare them with other theoretical settings. In order to achieve the accuracy required by the experimental information higher order radiative corrections to the decay \cite{Kuraev:2003gq} must be included and they have been implemented in our analysis. The numerical input for vector and axial-vector form factors is given in 
Tab.~\ref{tab:tablahadronic}.
\par
In the fourth column of Tab.~\ref{tab:tabladatos} the latest experimental data are given; in the fifth
and sixth we show the results provided by our analysis. We study the numerical impact of the 
momenta dependence of the form factors by setting the slopes to zero and we conclude that it is tiny~:
the $q^2$ dependence tends to increase the central value of $R$ but the modification is by far within
the errors.
The last two columns bring the results yielded by two- and three- flavour two-loop $\chi$PT calculations. The evaluation of the errors for the theoretical predictions is ruled by those in the form factors. The estimate of the latter has been done in the following way~: We assume no error coming from the slopes (since their numerical impact is very poor); to the vector form factor we assign the same error as that of the experimental determination $\sim$7\% and to the axial-vector form factor we attach the error of the experimental input. Finally the 
error given for the $\chi$PT calculations only considers the scale dependence that is a tiny 5\%.
\par
We conclude then that the corrections induced by the $q^2$ dependence of vector and axial-vector form factors
are numerically negligible unless the theoretical error is reduced. For this we would need a better determination of vector and, specially, axial-vector form factors at $q^2=0$. When comparing our results with experimental data, we see that our predictions are in agreement with previous estimates.
\par
As a final exercise we use the experimental data to fit the value of ${\cal T}= F_T f_T$ defined above. In order to reach this purpose we use the experimentally fitted values for the hadronic inputs $F_V(0)$
and $F_A(0)$, and our results for the slopes $\lambda_V$ and $\lambda_A$.
Finally, to extract the value of the $F_T$ coupling from the fit, we use our determination for the tensor form factor $f_T$. The value that we obtain is compatible with zero and its order of magnitude is compatible with that dictated by SUSY~:
\begin{equation}
F_T=(1\pm 14)\times 10^{-4} \, .
\end{equation}

\section{Conclusions}
Radiative pion decay has been a continuous source of debate between theoretical predictions and experimental
determinations. Nevertheless the latest analysis by the PIBETA Collaboration seems to bring a close agreement
between both sides. 
\par
In this article we have performed a detailed analysis of the structure-dependent amplitudes contributing
to $\pi \rightarrow e \, \nu_e \gamma$. The $q^2$ dependence of vector and axial-vector form factors, driven
by the Standard Model, has been rigorously constructed through the study of the $\braket{VVP}$ and 
$\braket{VAP}$ Green functions, by matching meromorphic {\em ans\"atze} with their leading OPE contributions. 
Moreover we have also required that our form factors are soft at high transfer of momenta. Hence we obtain
the most general (and simple) functions that satisfy all those constraints. The appropriate structure of 
the form factors requires a double vector resonance pole for the vector form factor and a single axial-vector
resonance pole for the axial-vector form factor. After a small momenta expansion
we compare our results with those of $\chi$PT and while in the vector sector we find complete
agreement, our slope for the axial-vector form factor is much larger than the one provided by
modelizations of local terms in the chiral framework.
\par
The role of a tensor contribution to the radiative pion decay has customarily been taken into account in order
to analyse the experimental results. We use those in order to fix the size of the contribution and we find 
that it is compatible with zero. Incidentally we have given a prediction for $f_V^{\perp}$ that
measures the coupling of a vector resonance $J^{PC}=1^{--}$ to the tensor current. Our results agree well with 
determinations from QCD sum rules and quenched lattice.
\par
We conclude that the Standard Model is able to embody the experimentally known features of the radiative pion
decay. As it happens with other decays involving non-perturbative strong effects, the rather large
size of the numerical incertitudes generated by our lack of knowledge of this QCD regime shows that this 
process is, at present, unsuitable for the search of New Physics.

\section*{Acknowledgments}
\begin{acknowledgement}
We thank M.~Bychkov and D.~Po\v cani\'c for useful discussions and for providing us with the 
latest experimental results on radiative pion decay. We also thank A.~Pich for a thorough reading
of the manuscript and his suggestions. The work of V.~Mateu is supported by a FPU contract (MEC).
This work has been supported in part by the EU
MRTN-CT-2006-035482 (FLAVIAnet), by MEC (Spain) under grant
FPA2004-00996 and by Generalitat Valenciana under grant GVACOMP2007-156.
\end{acknowledgement}

\appendix
\newcounter{vector}
\renewcommand{\thesection}{\Alph{vector}}
\renewcommand{\theequation}{\Alph{vector}.\arabic{equation}}
\setcounter{vector}{1}
\setcounter{equation}{0}
\section{$\braket{VVP}$ from a Lagrangian} \label{app:a}
We used the meromorphic {\em ansatz} in Eq.~(\ref{ansatz_2}) in order to determine the vector form 
factor in radiative pion decay. To reach the same result one could also proceed starting with a Lagrangian
like the one given by Resonance Chiral Theory (R$\chi$T).  This study, with one multiplet of vector 
resonances, was already carried out in Ref.~\cite{Ruiz-Femenia:2003hm}. However we have concluded 
that QCD constraints seem to indicate the need of including a second vector multiplet. In this appendix
we construct the short-distance constrained $\braket{VVP}$ function within R$\chi$T.
\par
In order to proceed we need to 
build the odd-intrinsic parity R$\chi$T Lagrangian with two multiplets of vector resonances. The
interaction pieces are~:
\begin{eqnarray}
{\cal L}_V^{\mathrm{even}} & = & \frac{F_V}{2 \sqrt{2}} \left\langle V_1^{\mu \nu} f_{+ \, \mu \nu} \right\rangle \, + \, 
\frac{F_V'}{2 \sqrt{2}} \left\langle V_2^{\mu \nu} f_{+ \, \mu \nu} \right\rangle  \, , \\
&& \nonumber \\
 {\cal L}_V^{\mathrm{odd}} & = & {\cal L}_{WZ} \, + \, 
i \, \varepsilon_{\mu \nu \alpha \beta} \left\{ \widetilde{C_7}^W \left\langle \chi_{-} f_+^{\mu \nu}
f_+^{\alpha \beta} \right\rangle \, \right. \nonumber \\
&& \; \; \; \; \; \; \; \; \;\; \;  \; \; \;\; \;  \; \; \;\; \;  \;\; \;  
\left. + \, i \, \widetilde{C_{22}}^W \left\langle \nabla_{\lambda} f_+^{\lambda \mu}
\left\{ f_+^{\alpha \beta}, u^{\nu} \right\} \right\rangle \right\} \, \nonumber \\
&&+ \sum_{i=1}^7 \frac{c_i}{M_{V_1}} \, {\cal O}_{V_1 JP}^i +
\sum_{i=1}^7 \frac{c_i'}{M_{V_2}} \, {\cal O}_{V_2 JP}^i \nonumber \\
&& + \sum_{i=1}^4 d_i \, {\cal O}_{V_1 V_1 P}^i + \sum_{i=1}^4 d_i'\,  {\cal O}_{V_2 V_2 P}^i 
\nonumber \\
&& + \sum_{n=a,b,c,d,e} d_n \, {\cal O}_{V_1 V_2 P}^n + d_f \, {\cal O}_{V_1 V_2 J}^f \, ,
\end{eqnarray}
where ${\cal L}_{WZ}$ is the Wess-Zumino Lagrangian \cite{Wess:1971yu} that arises at ${\cal O}(p^4)$ in 
$\chi$PT. As specified above only two operators contribute at ${\cal O}(p^6)$ \cite{Bijnens:2001bb}.
Operators ${\cal O}_{V_i JP}$ and ${\cal O}_{V_i V_i P}$ were already given in 
Ref.~\cite{Ruiz-Femenia:2003hm} 
and will not be repeated here. For the last part of the Lagrangian there are two subsets of 
pieces \cite{Pedrothesis}~:
\begin{itemize}
\item $V_1V_2P$ terms, which contain vertices with Goldstone and two vector resonances from different multiplets~:
\begin{eqnarray}
\mathcal{O}^a_{V_1V_2P}&=&\varepsilon_{\mu\nu\rho\sigma}
\left\langle \left\lbrace V_1^{\mu\nu},V_2^{\rho\alpha}\right\rbrace \nabla_\alpha u^\sigma\right\rangle 
\, , \nonumber\\
\mathcal{O}^b_{V_1V_2P}&=&\varepsilon_{\mu\nu\rho\sigma}
\left\langle \left\lbrace V_1^{\mu\alpha},V_2^{\rho\sigma}\right\rbrace \nabla_\alpha u^\nu\right\rangle 
\, , \nonumber\\
\mathcal{O}^c_{V_1V_2P}&=&\varepsilon_{\mu\nu\rho\sigma}
\left\langle \left\lbrace \nabla_\alpha V_1^{\mu\nu},V_2^{\rho\alpha}\right\rbrace  u^\sigma\right\rangle
\, , \nonumber \\
\mathcal{O}^d_{V_1V_2P}&=&\varepsilon_{\mu\nu\rho\sigma}
\left\langle \left\lbrace \nabla_\alpha V_1^{\mu\alpha},V_2^{\rho\sigma}\right\rbrace 
u^\nu\right\rangle 
\, , \nonumber \\
\mathcal{O}^e_{V_1V_2P}&=&\varepsilon_{\mu\nu\rho\sigma}
\left\langle \left\lbrace \nabla^\sigma V_1^{\mu\nu},V_2^{\rho\alpha}\right\rbrace 
 u_\alpha\right\rangle \, .
\end{eqnarray} 
\item $V_1 V_2 J$ terms, with two vector resonances from different multiplets and one pseudoscalar external source~:
\begin{equation}
\mathcal{O}^f_{V_1V_2J}=i\,\varepsilon_{\mu\nu\rho\sigma}
\left\langle \left\lbrace V_1^{\mu\nu},V_2^{\rho\sigma}\right\rbrace \chi_- \right\rangle \, .
\end{equation} 
\end{itemize} 
The result for the $\braket{VVP}$ Green function defined in Eq.~(\ref{eq:vvpgff}) is~:
\begin{eqnarray}
\Pi_{VVP}^{R\chi T}&=&- B_0 
\left\lbrace 
64 \, \widetilde{C_7}^W - 16 \, \widetilde{C_{22}}^W \, \frac{p^2+q^2}{r^2} \right. \nonumber \\
&&+\dfrac{Ar^2+B\left(p^2+q^2 \right) }{\left( M_{V_1}^2-p^2\right)
\left( M_{V_1}^2-q^2\right)r^2 }  \nonumber \\
&& +\, C \, \dfrac{1}{\left( M_{V_1}^2-p^2\right)
\left( M_{V_1}^2-q^2\right)}\, - \, \dfrac{N_C}{4\pi^2r^2} \nonumber \\
&&+ D\, \left( \dfrac{1}{M_{V_1}^2-p^2}+\dfrac{1}{M_{V_1}^2-q^2}\right) \nonumber \\ 
&&+\dfrac{1}{r^2}
\left( \dfrac{ Er^2+Kp^2+Gq^2}{M_{V_1}^2-p^2}+
\dfrac{ Er^2+Kq^2+Gp^2}{M_{V_1}^2-q^2}\right) \nonumber\\
&&+ \, \dfrac{A'r^2+B'\left(p^2+q^2 \right) }{\left( M_{V_2}^2-p^2\right)
\left( M_{V_2}^2-q^2\right)r^2 } \nonumber \\
&&+\, C'\,\dfrac{1}{\left( M_{V_2}^2-p^2\right)
\left( M_{V_2}^2-q^2\right)} \nonumber\\
&&+\,D'\,\left( \dfrac{1}{M_{V_2}^2-p^2}+\dfrac{1}{M_{V_2}^2-q^2}\right) \nonumber \\
&& +
\dfrac{1}{r^2}\left( \dfrac{ E'r^2+K'p^2+G'q^2}{M_{V_2}^2-p^2} \right. \nonumber \\
&& \left. \; \; \; \; \; \; \;  +\,  \dfrac{E'r^2+K'q^2+G'p^2}{M_{V_2}^2-q^2}\right)\nonumber\\
&&+ \dfrac{1}{r^2}\left[ \dfrac{A''r^2+B''p^2+Hq^2}{\left( M_{V_1}^2-p^2\right)
\left( M_{V_2}^2-q^2\right)} \right. \nonumber \\ 
&& \left.  \; \; \; \; \; \;\; + \,  \dfrac{A''r^2+B''q^2+Hp^2}{\left( M_{V_2}^2-p^2\right)
\left( M_{V_1}^2-q^2\right)}\right] \nonumber\\
&&+ \, C''\left[ \dfrac{1}{\left( M_{V_1}^2-p^2\right)\left( M_{V_2}^2-q^2\right)} \right. \nonumber \\
&& \left. \left. \; \; \; \; \; \; \, +\, 
\dfrac{1}{\left( M_{V_1}^2-q^2\right)\left( M_{V_2}^2-p^2\right)}\right] \right\rbrace \, ,
\end{eqnarray} 
where
\begin{eqnarray}
A&=&8 \,F_V^2\, \left(d_1-d_3 \right) \, , \nonumber \\
A'&=&8\, F_V'^2\, \left(d'_1-d'_3 \right) \, , \nonumber \\
B&=& 8\, F_V^2\, d_3 \, , \nonumber \\
B'&=& 8\, F_V'^2\, d'_3 \, ,\nonumber\\
C&=&64\, F_V^2\, d_2\, , \nonumber \\
C'&=&64\, F_V'^2\, d'_2\, ,\nonumber\\
D&=&-\dfrac{32\sqrt{2}\, F_V\, c_3}{M_{V_1}}\, , \nonumber \\
D'&=&-\dfrac{32\sqrt{2}\, F_V'\, c'_3}{M_{V_2}}\, ,\nonumber\\
E&=&-\dfrac{4\sqrt{2}\, F_V}{M_{V_1}}\, \left(c_1+c_2-c_5\right)\, , \nonumber \\
E'&=&-\dfrac{4\sqrt{2}\, F_V'}{M_{V_2}}\, \left(c'_1+c'_2-c'_5\right)\, ,\nonumber\\
K&=&-\dfrac{4\sqrt{2}\, F_V}{M_{V_1}}\, \left(-c_1+c_2+c_5-2c_6\right)\, , \nonumber \\
K'&=&-\dfrac{4\sqrt{2}\, F_V'}{M_{V_2}}\, \left(-c'_1+c'_2+c'_5-2c'_6\right)\, ,\nonumber\\
G&=&-\dfrac{4\sqrt{2}\, F_V}{M_{V_1}}\, \left(c_1-c_2+c_5\right)\, , \nonumber \\
G'&=&-\dfrac{4\sqrt{2}\, F_V'}{M_{V_2}}\, \left(c'_1-c'_2+c'_5\right)\,,\nonumber\\
C''&=&32\, F_V\, F_V'\, d_f\, , \nonumber \\
B''&=&4\, F_V\, F_V'\, \left(d_b+d_c-d_a-2d_d\right)\, , \nonumber\\
A''&=&4\, F_V\, F_V'\, \left(d_a+d_b-d_c\right)\, , \nonumber \\
H&=&4F_VF_V'\left(d_a+d_c-d_b\right) \, .
\end{eqnarray}
In terms of the parameters of our {\em ansatz} in Eq.~(\ref{ansatz_2}) we obtain~:
\begin{eqnarray}
c_{031}&=&-G-G'\,,\nonumber\\ 
c_{022}&=&-2(K+K')-\dfrac{N_C}{4\pi^2}\,, \nonumber \\
c_{121}&=&-D-D'-E-E'\,,\nonumber\\ 
c_{120}&=&(D+E)M_{V_2}^2+(D'+E')M_{V_1}^2\,,\nonumber\\
c_{111}&=&A+A'+2A''+C+C'+2C'' \nonumber \\
&& +2\left( M_{V_1}^2+M_{V_2}^2\right)(D+D'+E+E')\,, \nonumber\\
c_{021}&=&B+B'+B''+H+K\left( M_{V_1}^2+2M_{V_2}^2\right) \nonumber \\
&& +K'\left( 2M_{V_1}^2+M_{V_2}^2\right) \nonumber \\
&& +\left( M_{V_1}^2+M_{V_2}^2\right)\left(G+G'+\dfrac{N_C}{4\pi^2} \right) \,,\nonumber\\
c_{030}&=&GM_{V_2}^2+G'M_{V_1}^2\,,\nonumber\\
c_{110}&=&-(D+E)M_{V_2}^4-(D'+E')M_{V_1}^4 \nonumber \\
&& -(A+A''+C+C'')M_{V_2}^2 \nonumber \\
&& -(A'+A''+C'+C'')M_{V_1}^2\nonumber \\
&& -2(D+D'+E+E')M_{V_1}^2M_{V_2}^2\,,\nonumber \\
c_{011}&=&-2KM_{V_2}^4-2K'M_{V_1}^4-\left(M_{V_1}^2+M_{V_2}^2 \right)^2\dfrac{N_C}{4\pi^2} \nonumber \\
&& -2(H+B')M_{V_1}^2-2(B+B'')M_{V_2}^2\nonumber\\
&&-2(K+K'+G+G')M_{V_1}^2M_{V_2}^2 \,,\nonumber\\
c_{020}&=&-G'M_{V_1}^4-GM_{V_2}^4 \nonumber \\
&&-\left(K+K'+G+G'+\dfrac{N_C}{4\pi^2}\right)M_{V_1}^2M_{V_2}^2 \nonumber \\
&& -(B'+B'')M_{V_1}^2-(B+H)M_{V_2}^2\,,\nonumber\\
c_{010}&=&BM_{V_2}^4+B'M_{V_1}^4+(K'+G')M_{V_2}^2M_{V_1}^4 \nonumber \\
&& +(K+G)M_{V_1}^2M_{V_2}^4+ \nonumber\\
&&+(B''+H)M_{V_1}^2M_{V_2}^2+\left(M_{V_1}^2+M_{V_2}^2 \right)M_{V_1}^2M_{V_2}^2 \dfrac{N_C}{4\pi^2}\,,\nonumber\\
c_{100}&=&(A'+C')M_{V_1}^4+(A+C)M_{V_2}^4 \nonumber \\
&& +2(A''+C'')M_{V_1}^2M_{V_2}^2+\nonumber\\
&&+2(E'+D')M_{V_1}^4M_{V_2}^2+2(E+D)M_{V_2}^4M_{V_1}^2\,,\nonumber\\
c_{000}&=&-M_{V_2}^4M_{V_1}^4\dfrac{N_C}{4\pi^2}\, ,
\end{eqnarray} 
in units of $-B_0$.
Chiral symmetry, implemented in our Lagrangian, brings features that with the {\em ansatz} had
to be forced by hand. In this way we immediately find that $c_{300}=0$ and $c_{210}=0$. Moreover, 
as a bonus we also find $c_{200}=0$. 
\par
The rest of constraints are given in Eqs.~(\ref{eq:consvv},\ref{eq:consvv1}). In
addition we find five more relations~:
\begin{eqnarray}
\widetilde{C_7}^W &= & 0 \, , \nonumber \\
\widetilde{C_{22}}^W &= & 0 \, , \nonumber \\
 G + G' & = & 0 \, , \nonumber \\
D + D' + E + E' &=& 0 \, , \nonumber \\
2 \left( K + K' \right) & = & - \frac{N_C}{4 \pi^2} \, .
\end{eqnarray}
After applying all the constraints coming from the OPE expansion and Brodsky-Lepage asymptotic 
condition we obtain the following relations among the Lagrangian couplings~:
\begin{eqnarray}
4\, c_3+c_1 & = &  0\, , \nonumber \\
4\, c'_3+c'_1 & = &  0\, , \nonumber \\
c_1-c_2+c_5 & = & 0\, , \nonumber \\
c'_1-c'_2+c'_5 & = & 0 \, , \nonumber\\
&&  \nonumber \\
c_5 - c_6 + \frac{F_V' M_{V_1}}{F_V M_{V_2}} \left( c_5' - c_6' \right) & = &
\frac{M_{V_1}}{F_V} \, \frac{N_C}{64 \sqrt{2} \pi^2} \, , \nonumber \\
&& \nonumber \\
8 F_V^2 d_3+8 F_V'^2 d'_3+8F_VF_V'(d_c-d_d) &&\nonumber \\
+8\sqrt{2}F_V'\dfrac{M_{V_2}^2-M_{V_1}^2}{M_{V_2}}(c'_5-c'_6) && \nonumber \\
+ M_{V_1}^2\dfrac{N_C}{8\pi^2}& =& F^2\, ,\nonumber\\
&& \nonumber \\
4F_V^2 (d_1+8d_2)+4F_V'^2 (d'_1+8d'_2) && \nonumber \\
+4F_VF_V'(d_a+d_b-d_d+8d_f) && \nonumber \\
+4\sqrt{2}F_V'\dfrac{M_{V_2}^2-M_{V_1}^2}{M_{V_2}}(c'_5-c'_6) &&\nonumber \\+
M_{V_1}^2\dfrac{N_C}{16\pi^2}& = & F^2\, ,\nonumber\\
&&  \\
8M_{V_2}^2 F_V^2 d_3+8 F_V'^2M_{V_1}^2 d'_3 && \nonumber \\
+4F_VF_V'\left[M_{V_1}^2(d_b+d_c-d_a-2d_d) \right. && \nonumber \\
\left. +M_{V_2}^2(d_a+d_c-d_b) \right] &= &
-M_{V_1}^2M_{V_2}^2\dfrac{N_C}{8\pi^2}\, . \nonumber
\end{eqnarray} 

\appendix
\newcounter{vector1}
\renewcommand{\thesection}{\Alph{vector1}}
\renewcommand{\theequation}{\Alph{vector1}.\arabic{equation}}
\setcounter{vector1}{2}
\setcounter{equation}{0}
\section{Chiral Lagrangians with external tensor sources} \label{app:b}
In this appendix we determine the matrix element in Eq.~(\ref{eq:tensorhadron}) by incorporating
tensor sources to the resonance Lagrangian. The inclusion of those external fields in $\chi$PT has been
studied in Ref.~\cite{Cata:2007ns}.
\par
We extend the QCD Lagrangian to accommodate an external tensor source~:
\begin{eqnarray}
{\cal{L}}_{QCD}&=&{\cal{L}}^0_{QCD}\,+\,{\cal{L}}_{ext}\,,\label{QCD_extended}\\
{\cal{L}}_{ext}&=&{\bar{\psi}}\gamma_{\mu}(v^{\mu}+\gamma_5a^{\mu})\psi-{\bar{\psi}}(s-i\gamma_5p)\,\psi+{\bar{\psi}}\sigma_{\mu\nu}{\bar{t}}^{\mu\nu}\psi\,.\nonumber
\end{eqnarray}
The tensor source includes both octet and singlet currents
\begin{equation}
{\bar{t}}^{\mu\nu}=\sum_{a=0}^8\,\dfrac{\lambda^a}{2}\,{\bar{t}}^{\mu\nu}_a\,.
\end{equation} 
One finds that
\begin{equation}
{\bar{\psi}}\,\sigma_{\mu\nu}\,{\bar{t}}^{\mu\nu}\,\psi\,=\,{\bar{\psi}}_L\,\sigma^{\mu\nu}\,t^{\dagger}_{\mu\nu}\,\psi_R\,+\,{\bar{\psi}}_R\,\sigma^{\mu\nu}\,t_{\mu\nu}\,\psi_L\,,
\end{equation}
and the change of basis reads~:
\begin{eqnarray}\label{expr}
{\bar{t}}^{\mu\nu}&=&P_L^{\mu\nu\lambda\rho}\, t_{\lambda\rho}\,+\,P_R^{\mu\nu\lambda\rho}\, t^{\dagger}_{\lambda\rho}\, ,\nonumber\\
t^{\mu\nu}&=&P_L^{\mu\nu\lambda\rho}\,{\bar{t}}_{\lambda\rho}\,,
\end{eqnarray}
where $P_{L,R}^{\mu\nu\lambda\rho}$ are the chiral projectors for the tensor fields, given by
\begin{eqnarray}\label{projectors} 
P_R^{\mu\nu\lambda\rho}&=&\frac{1}{4}(g^{\mu\lambda}g^{\nu\rho}-g^{\nu\lambda}g^{\mu\rho}-i\varepsilon^{\mu\nu\lambda\rho})\, ,\nonumber\\
P_L^{\mu\nu\lambda\rho}&=&\left( P_R^{\mu\nu\lambda\rho}\right)^\dagger \,.
\end{eqnarray}
In order to mantain the chiral invariance of the extended QCD Lagrangian (\ref{QCD_extended}) the tensor source must transform as~:
\begin{equation}
t_{\mu\nu}\to g_R\,t_{\mu\nu}\,g_L^\dagger\,.
\end{equation}
It is convenient, in order to build an effective Lagrangian invariant under the chiral group,
to define the tensor operators~:
\begin{equation}
t_{\pm}^{\mu\nu}\,=\,u^{\dagger}\, t^{\mu\nu}\, u^{\dagger}\, \pm \, u\, t^{\mu\nu \, \dagger} \, u\, ,
\end{equation}  
transforming with the compensating field
\begin{equation}
 t_{\pm}^{\mu\nu}\to h(g,x)\,t_{\pm}^{\mu\nu}\,h(g,x)^\dagger\,.
\end{equation} 
To the lowest order in the chiral expansion we have one operator that contributes to both the radiative pion decay and the $\braket{VT}$ function,
\begin{equation}\label{O4}
{\cal{L}}_4^{\chi PT}\doteq\Lambda_1\langle t_+^{\mu\nu}f_{+\mu\nu}\rangle\, .
\end{equation}
When explicitly including resonances the corresponding interacting Lagrangian of interest reads~:
\begin{eqnarray}\label{RLag}
\mathcal{L}^{R\chi T} & \doteq & \widetilde{\Lambda_1}\langle t_+^{\mu\nu}f_{+\mu\nu}\rangle\, + \frac{F_{V}}{2\sqrt{2}}\langle V_{\mu\nu}f_{+}^{\mu\nu}\rangle \nonumber \\
&& +\sqrt{2}\,F_{VT}\,M_{V}\,\langle V_{\mu\nu}t_{+}^{\mu\nu}\rangle \, .
\end{eqnarray}
The couplings $F_V$ and $F_{VT}$ are related with those defined in Eq.~(\ref{eq:fvperp}) by
$f_V = \sqrt{2} \, F_V$ and $f_V^{\perp} = \sqrt{2} \, F_{VT}$.
Upon integration of the vector meson we can relate the couplings of the two Lagrangians~: 
\begin{equation} \label{mudep}
\Lambda_1= \widetilde{\Lambda_1} \, - \, \frac{F_{V}F_{VT}}{M_{V}} \, .
\end{equation} 
If resonance saturation of the chiral LECs would hold when including external tensor sources, $\Lambda_1$ 
should be given only by the second term in Eq.~(\ref{mudep}).  In fact this is the case when 
we enforce short-distance constraints. If we evaluate the $\braket{VT}$ Green function defined in 
Eq.~(\ref{eq:cond4}) we obtain~:
\begin{equation}
 \Pi_{VT}(p^2) \, = \, i \, \left( \widetilde{\Lambda_1} - \frac{F_V F_{VT} M_V}{M_V^2 - p^2} \right)\, , 
\end{equation}
that satisfies the OPE constraint in Eq.~(\ref{eq:cond4}) provided that~:
\begin{eqnarray}
 \widetilde{\Lambda_1} & = & 0 \, , \nonumber \\
F_V F_{VT}  & = & \frac{B_0 F^2}{M_V}\, . 
\end{eqnarray}
Using now the relation in Eq.~(\ref{eq:ftVT}) we get again the result in Eq.~(\ref{eq:tensorchiral}) for 
$f_T$.
\par
As mentioned in Section~\ref{sec:beyond} there are two form factors involved in the hadronic tensor matrix element
\begin{eqnarray}\label{eq:fgtff}
 &&\left\langle \gamma \left| \bar{u}\,\sigma_{\mu\nu}\,\gamma_5\, d\right|\pi^-\right\rangle =
\,-\,\dfrac{e}{2}\,f_T(q^2)\,(p_\mu\epsilon_\nu -p_\nu\epsilon_\mu)\\&&
\,-\,\dfrac{e}{2}\,g_T(q^2) \left[\, \epsilon\cdot q \,(p^\mu q^\nu-p^\nu q^\mu)+q\cdot p\,(q^\mu \epsilon^\nu-q^\nu \epsilon^\mu)\right] ,\nonumber
\end{eqnarray} 
where $p$ is the photon momentum and $q$ the transferred momentum by the tensor current.
To generate the second Lorentz structure one needs operators of higher order in the chiral expansion, such as 
\begin{equation}
Y_{93}=\left\langle \nabla_\mu t_+^{\mu\nu}\nabla^\alpha f_{+\alpha\nu}\right\rangle \,,
\end{equation}
of Ref.~\cite{Cata:2007ns}. In the framework of resonance chiral theory one needs in addition of the operators in (\ref{RLag}) the basis of odd-intrinsic parity operators of Ref.~\cite{Ruiz-Femenia:2003hm}. These new contributions do not modify result of the $f_T = f_T(0)$. The result reads~:
\begin{eqnarray}
f_T(q^2)&=&\dfrac{\sqrt{2}\,F_{VT}}{3\,F\,M_V}\left[2F_V 
 - \big( 4 \sqrt{2} \, M_V \, (c_1-c_2-c_5+2c_6) \right.  \nonumber \\
&& \; \; \;\; \; \;\; \; \; \; \; \; \; \; \left.  + 8 F_V \, d_3 \big) \, \dfrac{q^2}{M_V^2-q^2} \right] \, , \\ 
&& \nonumber \\
g_T(q^2)&=&\dfrac{8\,F_{VT}}{3\,F\,M_V^2}\left[ \frac{2\,M_V^2-q^2}{M_V^2-q^2} (c_1-c_2-c_5)\right. \,\nonumber\\
&&\; \; \; \; \; \; \; \;\; \; \; \; \; \; +\left. 2\,\frac{M_V^2}{M_V^2-q^2} \, c_6- 2\,c_7 \right.  \\
&& \; \; \; \; \; \; \; \;\; \; \; \; \; \;\left. + \sqrt{2} \frac{F_V}{M_V} 
\left( \frac{M_V^2}{M_V^2-q^2} \, d_3 +  d_4 \right) \right] \,. \nonumber 
\end{eqnarray} 
The spectral function of the tensor-tensor currents correlator to which the amplitude in Eq.~(\ref{eq:fgtff})
contributes behaves as a constant at high $q^2$ and leading order in $\alpha_S$ \cite{Craigie:1981jx}. Hence the $f_T(q^2)$ and $g_T(q^2)$ form factor should exhibit a smooth behaviour, vanishing at large transferred 
momentum. 
\par
Imposing that the $f_T(q^2)$ form factor vanishes at large momentum we get the constraint~:
\begin{equation}
 c_1-c_2-c_5+2 \,c_6\,= \, -\frac{\sqrt{2}}{4} \frac{F_V}{M_V} \left( 1 + 4 \,  d_3 \right) \, .
\end{equation} 
The same procedure with $g_T(q^2)$ gives~:
\begin{equation}
 c_1-c_2-c_5-2\, c_7 \, = \, - \sqrt{2} \, \frac{F_V}{M_V} \, d_4\, .
\end{equation}
 Interestingly enough these constraints fully determine both form factors~:
\begin{eqnarray}
f_T(q^2)\,&=&\,\dfrac{2\sqrt{2}\,F_{VT}\,F_V}{3\,F} \, \dfrac{M_V}{M_V^2-q^2}\,, \nonumber \\
g_T(q^2)\,&=&\,-\,\dfrac{f_T(q^2)}{M_V^2} \, .
\end{eqnarray}
Notice that the contribution of the $g_T(q^2)$ form factor to the matrix element under study
is fairly suppressed, typically ${\cal O}(q^2/M_V^2)$ over the $f_T(q^2)$ contribution.

\newcounter{vector2}
\renewcommand{\thesection}{\Alph{vector2}}
\renewcommand{\theequation}{\Alph{vector2}.\arabic{equation}}
\setcounter{vector2}{3}
\setcounter{equation}{0}
\section{LSZ formula for a soft pion} \label{app:c}
In this appendix we discuss the derivation of Eq.~(\ref{eq:LSZtwice}). Let us start defining three quark currents~:
\begin{eqnarray}
A_\pi^\mu(x)&=&\bar{d}(x)\,\gamma^\mu\,\gamma_5\,u(x)\,,\nonumber\\
T_5^{\mu\nu}(x)&=&\bar{u}(x)\,\sigma^{\mu\nu}\,\gamma_5\,d(x)\,,\\
T_\pi^{\mu\nu}&=&\bar{u}(x)\,\sigma^{\mu\nu}\,u(x)\,+\,\bar{d}(x)\,\sigma^{\mu\nu}\,d(x)\,.\nonumber
\end{eqnarray} 
Then we can construct the following GF that after partial integration can be expressed as~:
\begin{eqnarray}
&&\left\langle \gamma(p)\left|\,\partial_{\alpha} A^{\alpha}\;T_5^{\mu\nu}\right|0\right\rangle \,\equiv\\
&&i\,\int\mathrm{d}^4x\,e^{irx}
\left\langle \gamma(p)\left|T\left\lbrace \partial_\alpha A_\pi^\alpha(x),
T_5^{\mu\nu}(0)\right\rbrace \right|0\right\rangle \,=\,\nonumber\\
&& r_\alpha\int\mathrm{d}^4x\,e^{irx}\left\langle \gamma(p)\left|T\left\lbrace  A_\pi^\alpha(x),
T_5^{\mu\nu}(0)\right\rbrace \right|0\right\rangle\nonumber\\
&&-\,i\,\left\langle \gamma(p)\left|T_\pi^{\mu\nu}(0) \right|0\right\rangle\label{split}\,.
\end{eqnarray} 
We can relate this GF to the $\left\langle \gamma(p) \left| \bar{u}\,\sigma_{\mu\nu}\,\gamma_5\, d\right|\pi^-(r)\right\rangle$ matrix element (\ref{eq:snew_hort}) through the LSZ formula~: 
\begin{eqnarray}
&& \left\langle \gamma(p) \left| \bar{u}\,\sigma^{\mu\nu}\,\gamma_5\, d\right|\pi^-(r)\right\rangle=\nonumber\\
&&\lim_{r^2\to M_\pi^2}\dfrac{r^2-M_\pi^2}{\sqrt{2}\, F \, M_\pi^2}\,\left\langle \gamma(p)\left|\,\partial_{\alpha} A^{\alpha} \;T_5^{\mu\nu}\right|0\right\rangle\,.\label{LSZ_tensor}
\end{eqnarray} 
Then the $\left\langle \gamma(p)\left|\partial_{\alpha} A^{\alpha} \;T_5^{\mu\nu}\right|0\right\rangle$ GF must have a pole at $r^2\,=\,M_\pi^2$. Since the second piece in (\ref{split}) has no $r$-dependence it cannot have a pole, and this must come from the first term~:
\begin{eqnarray}
&&r_\alpha\,\int\mathrm{d}^4x\,e^{irx}\left\langle \gamma(p)\left|T\left\lbrace  A_\pi^\alpha(x),
T_5^{\mu\nu}(0)\right\rbrace \right|0\right\rangle\nonumber\\
&\equiv&\dfrac{1}{r^2-M_\pi^2}\,\mathcal{F}^{\mu\nu}(p,r)\,.
\end{eqnarray} 
As the divergence of the axial current is zero in the chiral limit, the left hand side must vanish for $M_\pi=0$, what implies~:
\begin{equation}
\left. \mathcal{F}^{\mu\nu}(p,r)\right|_{M_\pi=0}\,=\,i\,
\,r^2\,\left\langle \gamma(p)\left|T_\pi^{\mu\nu}(0) \right|0\right\rangle_{M_\pi=0}\,.
\end{equation} 
Finally, assuming that pion mass corrections are small, (\ref{LSZ_tensor}) yields the desired result~:
\begin{equation}
\left\langle \gamma(p) \left| \bar{u}\,\sigma^{\mu\nu}\,\gamma_5\, d\right|\pi^-(r)\right\rangle
\,=\,i\,\dfrac{1}{\sqrt{2}\,F}\,\left\langle \gamma(p)\left|T_\pi^{\mu\nu}(0) \right|0\right\rangle\, ,
\end{equation} 
that is also the result that stems from the soft pion theorem.


\begin{thebibliography}{10}

\bibitem{Weinberg:1978kz}
  S.~Weinberg,
  Physica A {\bf 96} (1979) 327;
  J.~Gasser and H.~Leutwyler,
  Annals Phys.\  {\bf 158} (1984) 142;
  J.~Gasser and H.~Leutwyler,
  Nucl.\ Phys.\  B {\bf 250} (1985) 465.

\bibitem{Holstein:1986uj}
  B.~R.~Holstein,
  Phys.\ Rev.\  D {\bf 33} (1986) 3316.

\bibitem{Bijnens:1996wm}
  J.~Bijnens and P.~Talavera,
  Nucl.\ Phys.\  B {\bf 489} (1997) 387
  [arXiv:hep-ph/9610269].

\bibitem{Ametller:1993hg}
  L.~Ametller, J.~Bijnens, A.~Bramon and F.~Cornet,
  Phys.\ Lett.\  B {\bf 303} (1993) 140
  [arXiv:hep-ph/9302219].

\bibitem{Geng:2003mt}
  C.~Q.~Geng, I.~L.~Ho and T.~H.~Wu,
  Nucl.\ Phys.\  B {\bf 684} (2004) 281
  [arXiv:hep-ph/0306165].

\bibitem{Ecker:1988te}
  G.~Ecker, J.~Gasser, A.~Pich and E.~de Rafael,
  Nucl.\ Phys.\  B {\bf 321} (1989) 311.

\bibitem{Cirigliano:2006hb}
  V.~Cirigliano, G.~Ecker, M.~Eidem\"uller, R.~Kaiser, A.~Pich and J.~Portol\'es,
  Nucl.\ Phys.\  B {\bf 753} (2006) 139
  [arXiv:hep-ph/0603205].

\bibitem{Kuhn:1990ad}
A.~Pich, \lq \lq QCD tests from Tau decay data``,
  SLAC Report 343, (1989), p. 416;
  J.~H.~K\"uhn and A.~Santamar\'{\i}a,
  Z.\ Phys.\  C {\bf 48} (1990) 445;
  F.~Guerrero and A.~Pich,
  Phys.\ Lett.\  B {\bf 412} (1997) 382
  [arXiv:hep-ph/9707347];
  M.~Jamin, A.~Pich and J.~Portol\'es,
  Phys.\ Lett.\  B {\bf 640} (2006) 176
  [arXiv:hep-ph/0605096].

\bibitem{GomezDumm:2003ku}
  D.~G\'omez Dumm, A.~Pich and J.~Portol\'es,
  Phys.\ Rev.\  D {\bf 69} (2004) 073002
  [arXiv:hep-ph/0312183].

\bibitem{Frlez:2003pe}
  E.~Frlez {\it et al.},
  Phys.\ Rev.\ Lett.\  {\bf 93} (2004) 181804
  [arXiv:hep-ex/0312029].

\bibitem{Pokanic06}
  D.~Po\v cani\'c, talk given at the 5$^{th}$ International Workshop on 
  Chiral Dynamics, Theory and Experiment \lq \lq Chiral Dynamics 2006``,
  September 18th-22nd (2006), Durham (USA).


\bibitem{Bolotov:1990yq}
  V.~N.~Bolotov {\it et al.},
  Phys.\ Lett.\  B {\bf 243} (1990) 308.

\bibitem{Poblaguev:1990tv}
  A.~A.~Poblaguev,
  Phys.\ Lett.\ B {\bf 238} (1990) 108.

\bibitem{Belyaev:1991gs}
  V.~M.~Belyaev and I.~I.~Kogan,
  Phys.\ Lett.\  B {\bf 280} (1992) 238;
  Yu.~Y.~Komachenko,
  Sov.\ J.\ Nucl.\ Phys.\  {\bf 55} (1992) 1384
  [Yad.\ Fiz.\  {\bf 55} (1992) 2487].

\bibitem{Herczeg:1994ur}
  P.~Herczeg,
  Phys.\ Rev.\  D {\bf 49} (1994) 247.


\bibitem{Poblaguev:2003ib}
  A.~A.~Poblaguev,
  Phys.\ Rev.\  D {\bf 68} (2003) 054020
  [arXiv:hep-ph/0307166].

\bibitem{Chizhov:2004tu}
 M.~V.~Chizhov,
  Mod.\ Phys.\ Lett.\  A {\bf 8} (1993) 2753
  [arXiv:hep-ph/0401217];
  M.~V.~Chizhov,
  arXiv:hep-ph/0310203;
  M.~V.~Chizhov,
  Phys.\ Part.\ Nucl.\ Lett.\  {\bf 2} (2005) 193
  [Pisma Fiz.\ Elem.\ Chast.\ Atom.\ Yadra {\bf 2N4} (2005) 7]
  [arXiv:hep-ph/0402105].

\bibitem{Quin:1993vh}
  P.~A.~Quin, T.~E.~Pickering, J.~E.~Schewe, P.~A.~Voytas and J.~Deutsch,
  Phys.\ Rev.\  D {\bf 47} (1993) 1247.

\bibitem{'tHooft:1973jz}
  G.~'t Hooft,
  Nucl.\ Phys.\  B {\bf 72} (1974) 461;
  E.~Witten,
  Nucl.\ Phys.\  B {\bf 160} (1979) 57.

\bibitem{Peris:1998nj}
  S.~Peris, M.~Perrottet and E.~de Rafael,
  JHEP {\bf 9805} (1998) 011
  [arXiv:hep-ph/9805442];
  M.~Knecht, S.~Peris, M.~Perrottet and E.~de Rafael,
  Phys.\ Rev.\ Lett.\  {\bf 83} (1999) 5230
  [arXiv:hep-ph/9908283];
  S.~Peris, B.~Phily and E.~de Rafael,
  Phys.\ Rev.\ Lett.\  {\bf 86} (2001) 14
  [arXiv:hep-ph/0007338].

\bibitem{Knecht:2001xc}
  M.~Knecht and A.~Nyffeler,
  Eur.\ Phys.\ J.\ C {\bf 21} (2001) 659
  [arXiv:hep-ph/0106034].

\bibitem{Ruiz-Femenia:2003hm}
  P.~D.~Ruiz-Femenia, A.~Pich and J.~Portol\'es,
  JHEP {\bf 0307} (2003) 003
  [arXiv:hep-ph/0306157].

\bibitem{Bijnens:2003rc}
  J.~Bijnens, E.~Gamiz, E.~Lipartia and J.~Prades,
  JHEP {\bf 0304} (2003) 055
  [arXiv:hep-ph/0304222].

\bibitem{Cirigliano:2004ue}
  V.~Cirigliano, G.~Ecker, M.~Eidem\"uller, A.~Pich and J.~Portol\'es,
  Phys.\ Lett.\  B {\bf 596} (2004) 96
  [arXiv:hep-ph/0404004].

\bibitem{Ecker:1989yg}
  G.~Ecker, J.~Gasser, H.~Leutwyler, A.~Pich and E.~de Rafael,
  Phys.\ Lett.\  B {\bf 223} (1989) 425.

\bibitem{otherGF}
  A. Pich, in Proceedings of the Phenomenology of Large $N_C$ QCD, 
  edited by R. Lebed (World Scientific, Singapore, 2002), p. 239, hep-ph/0205030;
  G.~Amor\'os, S.~Noguera and J.~Portol\'es,
  Eur.\ Phys.\ J.\  C {\bf 27} (2003) 243
  [arXiv:hep-ph/0109169];
  J.~Portol\'es and P.~D.~Ruiz-Femen\'{\i}a,
  Nucl.\ Phys.\ Proc.\ Suppl.\  {\bf 131} (2004) 170
  [arXiv:hep-ph/0311251];
  V.~Cirigliano, G.~Ecker, M.~Eidem\"uller, R.~Kaiser, A.~Pich and J.~Portol\'es,
  JHEP {\bf 0504} (2005) 006
  [arXiv:hep-ph/0503108];
  J.~Portol\'es,
  Nucl.\ Phys.\ Proc.\ Suppl.\  {\bf 164} (2007) 292
  [arXiv:hep-ph/0509279].

\bibitem{Bryman:1982et}
  D.~A.~Bryman, P.~Depommier and C.~Leroy,
  Phys.\ Rept.\  {\bf 88} (1982) 151.

\bibitem{Moussallam:1994xp}
  B.~Moussallam,
  Phys.\ Rev.\  D {\bf 51} (1995) 4939
  [arXiv:hep-ph/9407402].

\bibitem{Wess:1971yu}
  J.~Wess and B.~Zumino,
  Phys.\ Lett.\  B {\bf 37} (1971) 95;
  E.~Witten,
  Nucl.\ Phys.\  B {\bf 223} (1983) 433.

\bibitem{mateu-jamin-pich}
  V.~Mateu,
  arXiv:hep-ph/0610262;
V.~Mateu, M.~Jamin and A.~Pich, work in preparation.

\bibitem{Floratos:1978jb}
  E.~G.~Floratos, S.~Narison and E.~de Rafael,
  Nucl.\ Phys.\  B {\bf 155} (1979) 115.

\bibitem{Rosell:2006dt}
  I.~Rosell, J.~J.~Sanz-Cillero and A.~Pich,
  JHEP {\bf 0701} (2007) 039
  [arXiv:hep-ph/0610290];
  I.~Rosell,
  arXiv:hep-ph/0701248.

\bibitem{Lepage:1980fj}
  G.~P.~Lepage and S.~J.~Brodsky,
  Phys.\ Rev.\  D {\bf 22} (1980) 2157.

\bibitem{maxim}
  M.~A.~Bychkov, talk given at the APS April Meeting-2007, April 14th-17th (2007), Jacksonville (USA);
D.~Po\v cani\'c, talk given at INT, June 4th (2007), Seattle (USA).

\bibitem{Bijnens:2001bb}
  J.~Bijnens, L.~Girlanda and P.~Talavera,
  Eur.\ Phys.\ J.\ C {\bf 23} (2002) 539
  [arXiv:hep-ph/0110400].

\bibitem{Strandberg:2003zf}
  O.~Strandberg, Master of Science Thesis,
  ``Determination of the anomalous chiral coefficients of order $p^6$'' (2003),
  arXiv:hep-ph/0302064.

\bibitem{Bijnens:1989jb}
  J.~Bijnens, A.~Bramon and F.~Cornet,
  Z.\ Phys.\  C {\bf 46} (1990) 599.

\bibitem{Yao:2006px}
  W.~M.~R.~Yao {\it et al.}  [Particle Data Group],
  J.\ Phys.\ G {\bf 33} (2006) 1.

\bibitem{Ruiz-Femenia:2003gw}
  P.~D.~Ruiz-Femen\'{\i}a, A.~Pich and J.~Portol\'es,
  Nucl.\ Phys.\ Proc.\ Suppl.\  {\bf 133} (2004) 215
  [arXiv:hep-ph/0309345].

\bibitem{Pedrothesis}
  P.~D.~Ruiz-Femen\'{\i}a, \lq \lq Effective field theories for heavy and
light fermions``, PhD Thesis, University of Valencia (2003).


\bibitem{Cata:2007ns}
  O.~Cat\`a and V.~Mateu,
  arXiv:0705.2948 [hep-ph].


\bibitem{Craigie:1981jx}
  N.~S.~Craigie and J.~Stern,
  Phys.\ Rev.\  D {\bf 26} (1982) 2430.

\bibitem{Jamin:2002ev}
  M.~Jamin,
  Phys.\ Lett.\  B {\bf 538} (2002) 71
  [arXiv:hep-ph/0201174].

\bibitem{Ioffe:1983ju}
  B.~L.~Ioffe and A.~V.~Smilga,
  Nucl.\ Phys.\  B {\bf 232} (1984) 109.

\bibitem{He:1996wy}
  H.~x.~He and X.~D.~Ji,
  Phys.\ Rev.\  D {\bf 54} (1996) 6897
  [arXiv:hep-ph/9607408];
  L.~S.~Kisslinger,
  Phys.\ Rev.\  C {\bf 59} (1999) 3377
  [arXiv:hep-ph/9804320].

\bibitem{Ball:2002ps}
  P.~Ball and V.~M.~Braun,
  Phys.\ Rev.\  D {\bf 54} (1996) 2182
  [arXiv:hep-ph/9602323];
  P.~Ball, V.~M.~Braun and N.~Kivel,
  Nucl.\ Phys.\  B {\bf 649} (2003) 263
  [arXiv:hep-ph/0207307].

\bibitem{Bakulev:1999gf}
  A.~P.~Bakulev and S.~V.~Mikhailov,
  Eur.\ Phys.\ J.\  C {\bf 17} (2000) 129
  [arXiv:hep-ph/9908287].

\bibitem{Braun:2003jg}
  V.~M.~Braun, T.~Burch, C.~Gattringer, M.~Gockeler, G.~Lacagnina, S.~Schaefer and A.~Schafer,
  Phys.\ Rev.\ D {\bf 68} (2003) 054501
  [arXiv:hep-lat/0306006].

\bibitem{Becirevic:2003pn}
  D.~Becirevic, V.~Lubicz, F.~Mescia and C.~Tarantino,
  JHEP {\bf 0305} (2003) 007
  [arXiv:hep-lat/0301020].

\bibitem{Oscar-Vicent}
 O.~Cat\`a and V.~Mateu, work in preparation.


\bibitem{Kuraev:2003gq}
  E.~A.~Kuraev and Yu.~M.~Bystritsky,
  Phys.\ Rev.\ D {\bf 69} (2004) 114004
  [arXiv:hep-ph/0310275].

\end{thebibliography}
\end{document}